\documentclass[aps,twocolumn,superscriptaddress]{revtex4-1}
\usepackage{epsfig,graphicx,times}
\usepackage{amstext}
\usepackage{amsmath}
\usepackage{amssymb}
\usepackage{graphicx}
\usepackage{latexsym}
\usepackage{bm}
\usepackage[colorlinks,citecolor=blue, linkcolor=blue,hyperindex,CJKbookmarks,dvipdfm]{hyperref}
\newcommand{\Tr}{\text{Tr}}
\begin{document}
\title{Cram\'er-Rao bound and quantum parameter estimation with non-Hermitian systems }
\author{Jianning Li}
\affiliation{Center for Quantum Sciences and School of Physics, Northeast Normal University, Changchun 130024, China}
\author{ Haodi Liu }
\affiliation{Center for Quantum Sciences and School of Physics, Northeast Normal University, Changchun 130024, China}
\author{Zhihai Wang}
\affiliation{Center for Quantum Sciences and School of Physics, Northeast Normal University, Changchun 130024, China}
\author{Xuexi Yi}
\email{yixx@nenu.edu.cn}
\affiliation{Center for Quantum Sciences and School of Physics, Northeast Normal University, Changchun 130024, China}
\affiliation{Center for Advanced Optoelectronic Functional Materials Research, and Key Laboratory for UV-Emitting Materials and Technology of
Ministry of Education, Northeast Normal University, Changchun 130024, China}

\date{\today}

\begin{abstract}
The quantum Fisher information constrains the achievable precision in parameter estimation via the quantum Cram\'er-Rao bound, which has attracted much attention in Hermitian systems since the 60s of the last century. However, less attention has been paid to non-Hermitian systems.
In this Letter, working with different logarithmic operators, we derive two previously unknown expressions for quantum Fisher information, and two Cram\'er-Rao bounds  lower than the well-known one are found for non-Hermitian systems. These lower bounds are  due to the merit of non-Hermitian observable and it can be understood as a result of extended regimes of optimization. Two experimentally feasible examples are presented to illustrate the theory, saturation of these bounds and estimation precisions beyond the Heisenberg limit are predicted and discussed. A setup  to measure non-Hermitian observable is also proposed.
\end{abstract}


\maketitle
{\it Introduction.}---Quantum parameter estimation \cite{helstrom67,helstrom76,holevo82,degen17,pezze18}
aims at measuring the value of a continuous
parameter $\theta$ encoded in the state $\rho(\theta)$ of a quantum system. It gives better precision than the same measurement performed in a classical framework and plays a crucial role in
the advancement of physics. The estimation process generally consists of two steps. In the first step,  the state $\rho({\theta})$  is prepared and measured. A  simple way to prepare the state $\rho(\theta)$ is to evolve a reference initial state $\rho_0$ under a signal Hamiltonian $H_{\theta}$ that encodes the parameter $\theta$, $\rho_t(\theta)=e^{-iH_{\theta}t}\rho_0e^{iH_{\theta}t}.$ In the second step, the estimation  of $\theta$ is obtained  by data processing the measurement outcomes,  aiming at  the smallest  uncertainty $\Delta \theta$  given  finite resources such as time and number of particles. The uncertainty is bounded by the Cram\'er-Rao bound (CRB), which  expresses a lower  limit on the variance of unbiased estimations, stating that the variance of any such estimation is at least as high as the inverse of the Fisher information  \cite{fisher25,matson06}.

The quantum Cram\'er-Rao bound works in general with  Hermitian quantum systems \cite{frieden90,braunstein94,liu20} in order to meet the requirement of quantum mechanics. However,
recent studies found that non-Hermitian systems with unbroken PT symmetries  also possess a real spectrum \cite{bender98}. In fact, non-Hermiticity is ubiquitous in the quantum world \cite{bender07, guo09}, including systems with gain and/or loss \cite{konotop16m,feng17,ganainy19,miri19,ozdemir19}, many-body localization and   dynamical stability of non-Hermitian systems  \cite{hamazaki19}, as well as sensors designed with non-Hermitian systems \cite{berry04,heiss12}.

These facts  give rise to a question  that in non-Hermitian quantum systems  \cite{moiseyev11} what is the Cram\'er-Rao bound? This question was answered about 5 decades ago by Yuen and Lax in Ref. \cite{yuen73}, where the authors proposed an idea to estimate a complex parameter by measuring a non-Hermitian observable. Unfortunately, this theory was connected the complex parameter with the estimator by assuming that the average of the estimator is exactly the estimate of the parameter. This assumption would lead  to a flaw that the error propagation function is unity, resulting in a  less regime of non-Hermitian observables for optimization and experiments.  Consider that the study of non-Hermitian system and their unique properties have attracted fast growing interest in the last two decades, revisiting the  quantum Cram\'er bound and its consequent estimation theory with current technologies is highly desired for non-Hermitian systems. We should address that there are estimation protocols (or sensors) based on non-Hermitian system recently \cite{wiersig14,liu16,chen17,hodaei17,djorwe19,mao20}, but all analyses so far are based on either the Fisher information resulting from the  Hermitian  quantum Cram\'er-Rao bound, or the properties of the exceptional points.

In this Letter, we first develop an uncertainty relation for non-Hermitian operators, then we derive two previously unknown expressions  for quantum Fisher  information with different logarithmic derivatives. Two quantum Cram\'er-Rao bounds for non-Hermitian systems are  defined. We found that the Fisher information  is significantly increased due to the use of non-Hermitian operators, in particular for systems in mixed states. This is in contrary to the results of Fisher information with Hermitian symmetric logarithmic derivatives. Saturation of the two bounds is analyzed  and the optimal measurement to attain the bounds is derived. We elucidate the feature of non-Hermitian quantum  Fisher information   with GHZ states of trapped ions  \cite{pezze18,leibried05} and the phase estimation setup with Mach-Zehnder interferometer.  Comparison with the situation of non-Hermitian signal Hamiltonian is also presented and discussed.

{\it Non-Hermitian uncertainty relation and Quantum Cram\'er bound.}---
Higher estimation precision demands more resources. The trade-off between the precision and the resources required is determined  by the uncertainty principle, which constrains  to what extent complementary variables  maintain their averaged values and leads to the Heisenberg limit \cite{caves81, chin12, jarzyna15, giovannetti11,luis17,bai19} in quantum parameter estimation. One might wonder whether this is the case and how the uncertainty relation changes in non-Hermitian quantum mechanics \cite{moiseyev11}.

To explore the possible change of the uncertainty relation and introduce the variance for non-Hermitian systems, we introduce two  operators  $A$ and $B$, which are linear but not necessarily  Hermitian. Defining $\Delta A=A-\langle A\rangle$, $\Delta B=B-\langle B\rangle$ and  $O=\chi \Delta A+i\Delta B$  with $\chi$ a real parameter, $i^2=-1$ and $\langle Z\rangle$ being the expectation value  of operator $Z$,   we have
$\langle O^{\dagger}O\rangle\ge 0.$  Simple algebra yields \cite{supp},
$\langle\Delta A^\dagger \Delta A\rangle\langle\Delta B^\dagger \Delta B \rangle \geq \frac 1 4 |\langle C\rangle|^2,$
where $C$ was defined by $C=i(A^\dagger B-B^\dagger A) \equiv i[A, B]_a.$
Operator $C$ is a Hermitian operator regardless of $A$ and $B$ being Hermitian or not. $[X,Y]_a$ defines an abnormal commutation for operator $X$ and $Y$, which can be applied to determine whether a time-independent operator  is a constant of motion for systems governed by non-Hermitian Hamiltonian \cite{rivero20}.

We define $\sigma_A^2\equiv \langle\Delta A^\dagger \Delta A\rangle$ as the variance  for non-Hermitian operator $A$, which reduces to the traditional variance  when $A$ is Hermitian. This uncertainty relation as well as that we will present in the following also hold for unitary operators $A$ and $B$ \cite{bong18,yu19}.
In non-Hermitian quantum systems, the expectation value of dynamical variable $A$
might take complex values, $\langle A\rangle=|\langle A\rangle|e^{i\alpha}$.  The absolute value $|\langle A\rangle|$  and its phase $\alpha$
are both measurable quantities. For example, in scattering experiments the peaks in the
cross sections are obtained when the projectiles have energy which is equal to the absolute value of the energy, not the real part of the energy \cite{moiseyev11}. This complex  expectation  of  non-Hermitian operator  is equal to the weak value of the
positive-semidefinite part of the operator multiplied by a known complex
number \cite{pati15,nirala19,sahoo20}.

A stronger uncertainty relations for non-Hermitian operators $A$ and $B$ follows from the Schwarz inequality, $\langle F|F\rangle \langle G|G\rangle \geq |\langle F|G\rangle |^2$ with $|F\rangle=\Delta A|\Psi\rangle$ and $|G\rangle=\Delta B|\Psi\rangle$ and $|\Psi\rangle$ being an arbitrary  state of system,
\begin{eqnarray}\label{strong_sd}
\langle \Delta A^{\dagger}\Delta A\rangle \langle \Delta B^{\dagger} \Delta B\rangle \geq |\langle A^{\dagger}B\rangle -\langle A^{\dagger}\rangle \langle B\rangle|^2.
\end{eqnarray}
This is a slightly stronger inequality for the variance of operators $A$ and $B$, which together with error propagation \cite{supp}

\begin{equation}
(\Delta\theta)^2=\frac{\langle\Delta A^{\dagger}\Delta A\rangle}{\partial\langle A^\dagger\rangle/\partial \theta \cdot \partial\langle A\rangle/\partial \theta}
\end{equation}
leads to a quantum Cram\'er-Rao bound for non-Hermitian system,
\begin{equation}\label{bound1}
(\Delta\theta)^2\geq\frac{1}{F_{nH}^{(1)}},
\end{equation}
where $F_{nH}^{(1)}=\Tr (\rho L^{(1)\dagger} L^{(1)})$ is defined as non-Hermitian quantum Fisher information with \cite{helstrom73ijtp,supp}
\begin{equation}\label{LandLd}
\frac {\partial \rho}{\partial \theta}=\frac 1 2(L^{(1)}\rho+\rho L^{(1)\dagger}),
\end{equation}
where $\theta$ is the real parameter to be estimated based on density matrix $\rho(\theta)$ and Hermitian operator $A$.  The slightly stronger uncertainty relation Eq.(\ref{strong_sd}) reduces to  the Robertson-Schr\"odinger uncertainty relation when both $A$ and $B$ are Hermitian and returns to the unitary uncertainty relation \cite{bong18,yu19} with both $A$ and $B$ being unitary. Eq.(\ref{LandLd}) was derived with an assumption that $A^\dagger=A$, but there is no requirement for $L$. This indicates that Eq.(\ref{bound1}) holds for both Hermitian and non-Hermitian $L$. In fact, when $L$ is Hermitian, the result  reduces to the widely used quantum Fisher information (denoted by $F_H$ hereafter) in the literature, while non-Hermitian $L$ would lead to an enhanced precision for parameter estimations as shown below.

For non-Hermitian operator $A$, we obtain the other quantum Cram\'er-Rao bound leading to non-Hermitian  quantum information $F_{nH}^{(2)}=\Tr (\rho L^{(2)\dagger} L^{(2)}),$ which has the same expression as in Eq.(\ref{bound1}) but with $L^{(2)}$ instead of  $L^{(1)}$  \cite{helstrom73ijtp,supp},
\begin{equation}\label{LandL}
\frac {\partial \rho}{\partial \theta}=L^{(2)}\rho=\rho L^{(2)\dagger}.
\end{equation}
There are two fundamental differences   between Hermitian and non-Hermitian system in the quantum Cram\'er-Rao bounds defined by  $F_{nH}^{(1)}$ and $F_{nH}^{(2)}$,  which deserve to be outlined. Firstly, in  Fisher information $F_{nH}^{(1)}$, the observable $A$ is Hermitian but the so-call symmetric logarithmic derivative $L$ might be not. While both $A$ and $L$ in $F_{nH}^{(2)}$ are not Hermitian.
The state   $\rho(\theta)$ that encodes  the parameter  is Hermitian, however the signal Hamiltonian $H_\theta$ might not be Hermitian, i.e.,  $H_\theta\neq H_\theta^\dagger$. Secondly, in non-Hermitian system, the Fisher information is given  by $F_{nH}=\Tr (\rho L^{(y)\dagger} L^{(y)}),\, y=1, 2$ with $\frac {\partial \rho}{\partial \theta}=L^{(2)}\rho=\rho L^{(2)\dagger}$ and $\frac {\partial \rho}{\partial \theta}=\frac 1 2(L^{(1)}\rho+\rho L^{(1)\dagger})$ instead of $\frac {\partial \rho}{\partial \theta}=\frac 1 2 (L\rho+\rho L)$ in Hermitian system. As we will show later, $F_{nH}^{(1)}$ recovers the  well-known expression of quantum Fisher information $F_H$ for Hermitian system while  $F_{nH}^{(2)}$ can not. Whereas the Cram\'er-Rao bound defined by $F_{nH}^{(2)}$ can be saturated but that by $F_{nH}^{(1)}$ could not \cite{supp}.

{\it Fisher information of non-Hermitian system.}---
One of the main quests in  quantum parameter estimation is to find out the highest achievable precision with given resources and design schemes that attain that precision. In general, looking for the optimal resources and schemes is difficult since one needs to optimize over the input
state  that  encodes the parameter, the measurement that is performed at the output, and the estimator that assigns a parameter value to a given measurement outcome. One of the popular ways to obtain useful bounds in quantum parameter estimations, without the need for cumbersome
optimization, is to use
the quantum Fisher information. In the following, we will give an explicit expression for the non-Hermitian quantum Fisher information $F_{nH}^{(1)}$ and $F_{nH}^{(2)}$ in terms of the eigenstates and eigenvalues of the encoding density matrix $\rho(\theta)$.

Consider a $N$-dimensional quantum system.
The state of the system $\rho(\theta)$  is parameterized by the parameter
$\theta$  under estimation.  Without loss of generality, we assume the density matrix may not be of full rank and the spectral
decomposition of the density operator $\rho(\theta)$  is,
\begin{equation}
\rho(\theta)=\sum_i^M p_i(\theta)|\phi_i(\theta)\rangle\langle\phi_i(\theta)|, \,M\leq N.
\end{equation}
Here $p_i(\theta), i=1,2,...,M$ are required to be positive  for any value of $\theta$ and $\sum_i^M p_i(\theta)=1$, since $p_i(\theta)$ is the probability of the system in state $|\phi_i(\theta)\rangle$. The following normalization conditions $\langle\phi_i(\theta)|\phi_j(\theta)\rangle=\delta_{ij}$ for all $i$ and  $j$ are imposed. From the definition of  Fisher information, we can find that,
$$F_{nH}^{(y)}=\sum_{i=1}^M\sum_{j=1}^N p_i(L^{(y)\dagger})_{ij}L^{(y)}_{ji},\, y=1, 2.$$
Noticing the derivation  for the two Fisher information  $F_{nH}^{(1)}$ and $F_{nH}^{(2)}$ is different, we discuss it in the following separately.

We start our derivation of $F_{nH}^{(1)}$ by considering a special case that $L^{(1)\dagger}=e^{i\beta}L^{(1)}$ with $\beta$ a real parameter. This assumption together with
Eq.(\ref{LandLd}) leads to,
\begin{equation}\label{L1}
L^{(1)}_{ij}=\langle\phi_i|L^{(1)}|\phi_i\rangle=\frac{2(\partial_\theta \rho)_{ij}}{p_j+p_ie^{i\beta}}
\end{equation}
and $L^{(1)\dagger}_{ij}=e^{i\beta}L^{(1)}_{ij}.$ A notation $(\partial_\theta \rho)_{ij}=\left( \frac{\partial \rho(\theta)}{\partial \theta}\right)_{ij}$ was used.
Clearly, $L^{(1)}\neq L^{(1)\dagger}$ as we expected. It is worth noticing that
$L^{(1)}_{ij}$ is in principle supported by the full Hilbert space, not only that spanned by the
eigenstates of the density matrix.
Thus the value of $L^{(1)}_{ij}$ for $i, j > N$ can not be established by the  above
equations. However,  these terms play an important role in the quantum Fisher information.
This problem can be solved by  the use of completeness relation, $\sum_{i=1}^N|\phi_i\rangle\langle\phi_i|=1$, i.e.,  $\sum_{i=M+1}^N|\phi_i\rangle\langle\phi_i|=1-\sum_{i=1}^M|\phi_i\rangle\langle\phi_i|$. We finally  arrive at \cite{supp},
\begin{widetext}
\begin{eqnarray}
F_{nH}^{(1)}=\sum_{i=1}^M \frac{2(\partial_\theta p_i)^2}{p_i(1+\cos\beta)}+\sum_{i=1}^M 4p_i\langle \partial_\theta\phi_i|\partial_\theta\phi_i\rangle
+\sum_{i=1}^M\sum_{j=1}^M \left( \frac{4p_i(p_i-p_j)^2}{p_i^2+p_j^2+2p_ip_j\cos\beta}-4p_i\right)\langle\partial_\theta\phi_i|\phi_j\rangle
\langle\phi_j|\partial_\theta\phi_i\rangle,
\end{eqnarray}
\end{widetext}
where $|\partial_\theta\phi\rangle\equiv\frac{\partial |\phi\rangle}{\partial\theta}.$ This is one of the main results of this Letter. For pure state,
$F_{nH}^{(1)}$ reduce to $ F_{nH}^{(1)}=F_H= 4\langle \partial_\theta\phi|\partial_\theta\phi\rangle
-4\langle\partial_\theta\phi|\phi\rangle
\langle\phi|\partial_\theta\phi\rangle.$ This is the widely used quantum fisher information in literature. For mixed states with $\beta\rightarrow 0$, $F_{nH}^{(1)}$ reduces to the Hermitian Fisher information, whereas for mixed states with $\beta\rightarrow \pi$,
\begin{equation}
F_{nH}^{(1)}\simeq \sum_{i=1}^M 4p_i\langle \partial_\theta\phi_i|\partial_\theta\phi_i\rangle+F_c,
\end{equation}
where $F_c=\sum_{i=1}^M\frac{2(\partial_\theta p_i)^2}{p_i(1+\cos\beta)}|_{\beta\rightarrow \pi}.$
$F_c$ approaches to infinity in the limit of $\beta\rightarrow \pi$. This is not physical,  since $L_{ij}$ in Eq. (\ref{L1}) can not be established  when $\beta\rightarrow \pi$ and $i=j$. In the following discussion, we will focus on the case that $p_i$ are $\theta$-independent, such that $F_c=0.$

The quantum information $F_{nH}^{(2)}$ can be calculated by the same  technique. Different from the case of  $F_{nH}^{(1)}$, here
$(L^{(2)\dagger})_{ij}=\frac{(\partial_{\theta}\rho)_{ij}}{p_i}$, $L^{(2)}_{ji}=\frac{(\partial_{\theta}\rho)_{ji}}{p_i}$ for $p_i\neq 0.$   Although both $A$ and $L$ are not Hermitian, the Fisher information is  real and  given by \cite{supp},
\begin{eqnarray}
F_{nH}^{(2)}&=&\sum_{i=1}^M\frac{(\partial_\theta p_i)^2}{p_i}+\sum_{i=1}^M p_i\langle\partial_\theta\phi_i|\partial_\theta\phi_i\rangle\\
&+&\sum_{i=1}^M\sum_{j=1}^M\frac{p_j(p_j-2p_i)}{p_i}\langle\partial_\theta
\phi_i|\phi_j\rangle\langle\phi_j|\partial_\theta\phi_i\rangle.\nonumber
\end{eqnarray}
This is the other main result of this Letter. The quantum Fisher information $F_{nH}^{(2)}$ for pure state $|\phi\rangle$ reduces to
\begin{equation}\label{Fpure}
F_{nH}^{(2)}=\langle\partial_\theta\phi|\partial_\theta \phi\rangle-\langle \partial_\theta\phi|\phi\rangle
\langle\phi|\partial_\theta\phi\rangle,
\end{equation}
which is $1/4$ times smaller than $F_H$ in Hermitian system for pure stats. However, this is not the case for mixed states as shown in the examples below. The  saturation of these bounds and the experimental measurement of non-Hermitian operator will be discussed briefly at the end of this Letter.

{\it Examples.}---
Considering the Hermitian quantum Fisher information $F_H$  and the non-Hermitian quantum information $F_{nH}^{(1)}$ and $F_{nH}^{(2)}$ are almost same for pure stats, i.e., $F_H=F_{nH}^{(1)}$ and $F_H=4F_{nH}^{(2)}$ for pure states, we focus on the case of mixed states in the undergoing examples.
To simplify the expression, we consider a type of simplest mixed states---it is constructed  by only two orthogonal pure states with weight  $p_1$ and $p_2,$ and $p_1+p_2=1.$

We first consider  the  maximally entangled states (or Schr\"odinger cat state)
$|\phi_{1,2}\rangle=\frac {1} {\sqrt 2} (|a\rangle^{\otimes N}\pm|b\rangle^{\otimes N})$
as the input state, which has been created with up to $N=6$  $^9Be^+$ ions \cite{leibried05}
and $N=14$ $^{40}Ca^+$ ions \cite{monz11} in a linear
Paul trap. The encoding operator is a spin rotation defined by $U(\theta)=e^{-i\theta  J_z}$ with Hermitian signal Hamiltonian  $J_z=\sum_{j=1}^N s_z^{(j)}$. The case of non-Hermitian signal Hamiltonian will be discussed  at the end of examples. The states that encode the parameter $\theta$  to construct the mixed states are,
$$|\phi_{1,2}(\theta)\rangle=\frac {1} {\sqrt 2} (e^{\frac i 2 N\theta}|a\rangle^{\otimes N}\pm e^{-\frac i 2 N\theta}|b\rangle^{\otimes N}).$$
The mixed state is $\rho=\sum_{i=1}^2 p_i|\phi_i(\theta)\rangle\langle \phi_i(\theta)|.$
Notice that $p_i,\, i=1,2$ are $\theta$-independent.
The rotation $U(\theta)=e^{-i\theta  J_z}$ generates a  relative phase $N\theta$ between states $|a\rangle^{\otimes N}$ and $|b\rangle^{\otimes N}$.
Straightforward calculation gives the Hermitian Fisher information $F_H$ and the non-Hermitian Fisher information $F_{nH}^{(1)}$ and  $F_{nH}^{(2)}$ (setting $p_1=p$ and $\beta\rightarrow \pi$) \cite{supp},
\begin{eqnarray}
F_H&=&\left(1-4p(1-p)\right)N^2,\nonumber\\
F_{nH}^{(1)}&=&N^2,\quad F_{nH}^{(2)}=\frac{(2p-1)^2}{4p(1-p)}N^2.
\end{eqnarray}
The results are shown in Fig.\ref{qcbE1f1}. We find that the non-Hermitian Fisher information are always larger than the Hermitian Fisher information, except the points $p=0,0.5,1.$
\begin{figure}
\includegraphics*[width=0.7\columnwidth,
height=0.55\columnwidth]{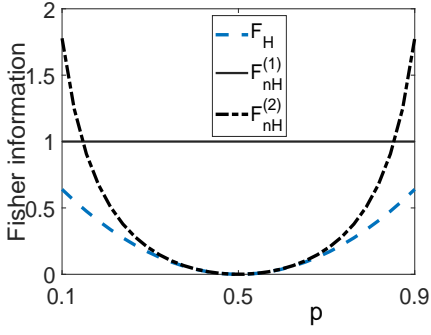} \caption{Quantum Fisher information $F_H$, $F_{nH}^{(1)}$ and  $F_{nH}^{(2)}$. The Fisher information was plotted in units of $N^2$. $p$ is the eigenvalue of the density matrix.} \label{qcbE1f1}
\end{figure}
It is worth addressing that at points $p=0,1$, the state is pure, so the Fisher information should be calculated by the formula of pure states, i.e., $F_H=F_{nH}^{(1)}=N^2$, and $F_{nH}^{(2)}=0.25N^2.$ As $p$ approach to $1$ and $0$, $F_{nH}^{(2)}$ tends to infinity, manifesting itself as a witness of transition from mixed states to pure states. The bound defined by $F_{nH}^{(1)}$ and $F_{nH}^{(2)}$ can be saturated by carefully chosen measurements. For details, see Supplemental Material \cite{supp}.

The second  example we will show is the phase estimation interferometric
schemes \cite{caves81,jarzyna12}, which works  with coherent and squeezed
vacuum states interfered at the beam-splitter of the Mach-Zehnder interferometer.
Let $a$ and $b$ be the anihilation operators of the two modes. $|\alpha\rangle=\exp(\alpha b^\dagger-\alpha^*b)|0\rangle_b$ is a coherent state of mode $b$ and  $|r\rangle=\exp[\frac 1 2 r^* a^2-\frac 1 2 r (a^\dagger)^2]|0\rangle_a$ is a squeezed vacuum state of mode $a$ with squeezing parameter $r$. After the input state \cite{supp}   has
evolved through a  beam splitter defined  by $B_\pi=\exp[-\frac i 2\pi(ab^\dagger+a^\dagger b)]$, the estimated parameter $\theta$ is encoded into the state by  relative phase shift operator $U(\theta)=\exp[-i\theta H],$ where $H$ is the signal Hamiltonian  of mode $a$, $H=a^{\dagger}a$. For the other beam splitter, see Supplemental Materials \cite{supp}.  We consider the following states that encode the parameter $\theta$, $\rho=\sum_{i=1}^2 p_i|\phi_i(\theta)\rangle\langle \phi_i(\theta)|$ with
\begin{eqnarray}
|\phi_{1,2}(\theta)\rangle&=&U(\theta)BS(r)D(\alpha)|0\rangle_a\otimes |X_{1,2}\rangle_b,
\end{eqnarray}
where $|X_1\rangle_b=|0\rangle_b$, $|X_2\rangle_b=b^{\dagger}|0\rangle_b,$ and $p_1+p_2=1.$
Tedious but straightforward calculations show that (set $p_1=p$)  \cite{supp},
\begin{eqnarray}
F_H&=&4|\alpha|^2(4p^2-6p+3), \nonumber\\
F_{nH}^{(1)}&=&4|\alpha|^4+(20-16p)|\alpha|^2+4(1-p),\nonumber\\
F_{nH}^{(2)}&=&\frac{2p^3-2p^2+1}{p(1-p)}|\alpha|^2.
\end{eqnarray}
\begin{figure}
\includegraphics*[width=0.8\columnwidth,
height=0.6\columnwidth]{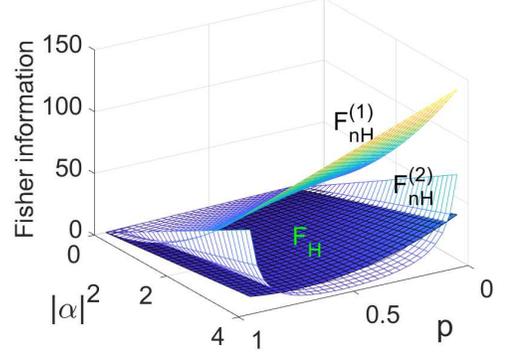} \caption{Quantum information as a function of photon number $|\alpha|^2$ and the eigenvalue of density matrix $p$.} \label{qcbE2f1}
\end{figure}
We have performed numerical calculations for the Fisher information. The results are shown in Fig. \ref{qcbE2f1}. We find that the two non-Hermitian quantum Fisher information favor $p\sim 0$. This can be understood as that squeezed Fock states benefit the parameter estimation more than the squeezed vacuum state.  $F_{nH}^{(1)}$ with term $|\alpha|^4$  bounds the estimation precision beyond the Heisenberg limit of $|\alpha|^2$.

One might wonder what is the difference between the non-Hermitian Fisher information and the Hermitian Fisher information with non-Hermitian signal Hamiltonian. To answer this question, we replace the Hermitian operator $J_z$ by $J_z(1-i\gamma)$ \cite{supp} in the first example to calculate the Hermitian Fisher information $F_H$, and the result will be denoted by $F_{H}^{nhs}$. We  compare $F_{H}^{nhs}$ with the  Hermitian  Fisher information $F_H$ with Hermitian signal Hamiltonian $J_z$. In other words, the comparison is conducted between the Hermitian Fisher information with Hermitian signal and non-Hermitian signal Hamiltonian for pure states. This comparison together with what we had in the two examples would show the difference between the Hermitian  and non-Hermitian Fisher information.
Calculation \cite{supp} finds,
\begin{eqnarray}
F_H&=&N^2,\nonumber\\
F_{H}^{nhs}&=&{|\bar N|^2}\left[\frac{}{}1-\tanh^2(\gamma N\theta)\right],
\end{eqnarray}
where  $\bar N=N(1-i\gamma).$
Numerical results are given in Fig.\ref{qcbE3f1}. We find $F_{H}^{nhs}$ is smaller than $F_H$  for almost all $\gamma$ and $\theta$, except  $\gamma$ and $\theta$  very close to zero.
The difference between  $F_{H}^{nhs}$ and  $F_H$ can be interpreted by
a residual dependence of the state on the parameter under estimation  apart from the statistical manifold of states.  This happens, for example, when the eigenstates of the density matrix
have a parameter-dependent normalization \cite{seveso17}.
\begin{figure}
\includegraphics*[width=0.8\columnwidth,
height=0.6\columnwidth]{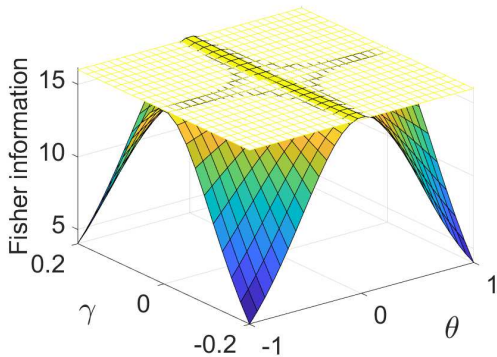} \caption{Hermitian quantum Fisher information $F_H$ (yellow-mesh) and  the non-Hermitian Fisher information $F_{H}^{nhs}$(blue-surf) as a function of estimation parameter $\theta$ and the loss/gain  rate $\gamma$. $\theta$ is chosen  in units of $\pi$, and  $N=4.$ } \label{qcbE3f1}
\end{figure}

In the second example, we take $\bar H=H(1-i\gamma)$ instead of $H$  as the signal Hamiltonian to account the effect of decoherence. $F_{H}^{nhs}$ is calculated with $\bar H$, while the Hermitian Fisher information $F_H$ is calculated with $H$,
\begin{eqnarray}
F_H&=&4|\alpha|^2,\nonumber\\
F_{H}^{nhs}&=&4|\alpha|^2(1+\gamma^2)e^{-2\gamma\theta}.
\end{eqnarray}
The scaling of Fisher information with the particle number remains unchanged, indicating there is no significant enhancement to parameter estimation in this example.

{\it Saturation of the bounds and feasible experiments to measure an non-Hermitian variable.}--- To saturate the bounds, the optimal measurement $A^{opt}$ has to meet $A^{opt}=\gamma L$ and $(A^{opt})^\dagger L=L^\dagger A^{opt}$ simultaneously \cite{supp}. These requirements can not be satisfied  for bound  $1/F_{nH}^{(1)}$, as $A$ is Hermitian but $L$ is not as we stated earlier. Even if we lift the constrain on the Hermiticity of $A$, the bound $1/F_{nH}^{(1)}$ can not be saturated too due to the requirement of $\beta=\pi$. This claim is confirmed by our numerical simulation, see Supplemental Material \cite{supp}. The situation is different for bound $1/F_{nH}^{(2)}$. Simple analyses show that the measurement variable
\begin{eqnarray}\label{Aopt}
A^{opt}_{ij}\equiv \langle \phi_i|A^{opt}|\phi_j\rangle=\gamma \frac{(\partial_{\theta}\rho)_{ij}}{p_j}
\end{eqnarray}
with real $\gamma$ can saturate the bound \cite{supp}. For the example with trapped ions, $A^{opt}$ can be measure in a single-shot experiment in an interferometer setup \cite{supp}.

{\it Conclusions.}---The framework of quantum mechanics in which observable are associated with Hermitian operators is too narrow to discuss parameter estimation. Considering in the past two decades the non-Hermitian physics has attracted fast growing interest in various field of research, we first derived an uncertainty relation for non-Hermitian operators, then we deduce a previously unknown expression for non-Hermitian Fisher information.  Two Cram\'er-Rao bounds that in some cases one of them,  and sometimes both of them, are superior to the previous result are found. The saturation of these bounds is  analysed and the optimal measurement to attain the bounds are given. The theory was illustrated with  two experimentally feasible systems.  The setup to measure non-Hermitian observables is also proposed.

We thank Xiaoming Lu for helpful discussions. This work was supported by the National Natural Science Foundation of China (NSFC) under Grants No. 11775048, and No. 12047566.

\onecolumngrid
\clearpage

\begin{center}
\textbf{\Large Supplementary material for: \\}
	\textbf{\large Cram\'er-Rao bound and quantum parameter estimation with non-Hermitian systems}
\end{center}

\setcounter{equation}{0} \setcounter{figure}{0} 
\makeatletter

\renewcommand{\theequation}{S\arabic{equation}}
\renewcommand{\thefigure}{S\arabic{figure}}
\renewcommand{\thesection}{S\arabic{section}}

This supplemental material provides detailed  derivations and  calculations for the results in the main text. We give numbers to the equations and figures here with "S" in contrast with that in the main text, for example, Eq.(S1), Fig. S1. Numbers without "S" refer to the items in the main text, e.g., Eq. (1), Fig. 1. This materials are organized  as follows.
\begin{itemize}
\item[S1.] We derive two Cram\'e-Rao bounds from uncertainty relations of non-Hermitian operators.
\item[S2.] Non-Hermitian Fisher information is defined and previously unknown expressions to calculate the Fisher information is given.
\item[S3.] Two examples to illustrate the non-Hermitian Fisher information are presented.
\item[S4.] We present discussion and comparison between the non-Hermitian Fisher information and the Hermitian Fisher information with non-Hermitian signal Hamiltonian.
\item[S5.] We discuss the second example with the other beam splitter.
\item[S6.] This section devotes to discussion of  saturation of Cram\'e-Rao bounds defined by the non-Hermitian Fisher information.
\item[S7.] We focus on the measurement of non-Hermitian operators. Detailed experimental proposal is given.
\end{itemize}

\maketitle

\section{S1.\, Uncertainly relation and Cram\'e-Rao bound for non-Hermitian system} \label{nHcrb}
The precision of  parameter estimation is limited by the quantum Cram\'er-Rao bound (CRB), which was an extension of  Cram\'er-Rao bound  in classical statistics to quantum metrology. As an extension of classical method, CRB has a wide range of applications, meanwhile it lacks a direct physical picture. In this section, we would derive the quantum CRB from the uncertainty relations that root deeply in the quantum theory from the beginning of the last century.

Consider two linear operators $A$, $B$ and a real parameter $\chi$, for any state $|\Psi\rangle$, the following relation holds,
\begin{equation}
\langle\Psi | (\chi A^\dagger-i B^\dagger)(\chi A+i B)|\Psi\rangle \geq 0.
\end{equation}
Simple algebra yields,
\begin{equation}
\chi^2\langle\Psi |A^\dagger A|\Psi\rangle +\chi\langle\Psi|i(A^\dagger B-B^\dagger A)|\Psi\rangle +\langle \Psi| B^\dagger B|\Psi\rangle \geq 0.
\end{equation}
Noticing $\langle\Psi |A^\dagger A|\Psi\rangle\geq 0$ regardless of $A$ being Hermitian or not, and
$\langle\Psi|i(A^\dagger B-B^\dagger A)|\Psi\rangle$ is real as $C\equiv i(A^\dagger B-B^\dagger A)$ is Hermitian, we obtain
\begin{equation}
\langle\Psi| A^\dagger A|\Psi\rangle\langle \Psi| B^\dagger  B|\Psi \rangle \geq \frac 1 4 |\langle \Psi| C|\Psi\rangle|^2.
\end{equation}
Define $\Delta A=A-\langle \Psi|A|\Psi\rangle$ and the same for $\Delta B$, and note the commutation relation,
$[A,B]=[\Delta A, \Delta B],$ we have
\begin{equation}\label{ucr1}
\langle\Psi|\Delta A^\dagger \Delta A|\Psi\rangle\langle\Psi|\Delta B^\dagger \Delta B |\Psi \rangle \geq \frac 1 4 |\langle\Psi| C|\Psi\rangle|^2.
\end{equation}
This is the equation (not numbered) in the main text.

Although the proof is conducted for pure states, the uncertainty relation Eq.(\ref{ucr1}) holds  for mixed states. We prove this by introducing an ancilla $a$, such that a mixed state
$\rho=\sum_j q_j|\psi_j\rangle\langle \psi_j|$ can be purified to be
\begin{equation}
|\Psi^\prime\rangle=\sum_j\sqrt{q_j}|\psi_j\rangle\otimes|\phi_j\rangle_a,
\end{equation}
and the state of the system is obtained by tracing $|\Psi^\prime\rangle$ over the ancilla,  $\rho=\Tr_a|\Psi^\prime\rangle\langle\Psi^\prime|.$
With this consideration, Eq. (\ref{ucr1}) can be straightforwardly  extended to  the composite system consisting of the system and  the ancilla,
\begin{equation}\label{ucr2}
\langle\Psi^\prime|\Delta A^\dagger\otimes I_a \Delta A\otimes I_a |\Psi^\prime\rangle\langle\Psi^\prime|\Delta B^\dagger \otimes I_a \Delta B\otimes I_a|\Psi^\prime \rangle \geq \frac 1 4 |\langle\Psi^\prime| C\otimes I_a|\Psi^\prime\rangle|^2.
\end{equation}
Here, $I_a$ is the identity operator of ancilla $a$. Noticing $\langle\Psi^\prime|\Delta A^\dagger\otimes I_a \Delta A\otimes I_a |\Psi^\prime\rangle=\Tr(\rho \Delta A^\dagger \Delta A)$ and denoting $\Tr(\rho \Delta A^\dagger \Delta A)=\langle \Delta A^\dagger \Delta A \rangle$ with $\Tr$ representing the trace over the system, we finish the proof of the weaker uncertainly relation  in the main text.
The slightly stronger uncertainty relation in Eq.(1) with mixed states can be proved in a similar way.

In order to develop a Cram\'er-Rao bound for non-Hermitian systems, we first have to extend error propagation from Hermitian to non-Hermitian systems, see Eq.(2).
Assume a quantum state $\rho=\rho(\theta)$ depends on parameter $\theta$, the expectation value of operator
$A^\dagger$ and its conjugate  $A$ would then depend on the parameter. Let us
discretize the parameter $\theta$  by  $\theta_1, \theta_2, \theta_3,....$
The fluctuation is then,
\begin{eqnarray}
\langle \Delta A^\dagger\Delta A\rangle=\sum_j p_j\left[\langle A^\dagger\rangle (\theta_j)-\langle A^\dagger\rangle (\bar\theta)\right]\left[\langle A\rangle (\theta_j)-\langle A\rangle (\bar\theta)\right],
\end{eqnarray}
where $\bar\theta=\sum_j p_j\theta_j$, and $p_j$ ($j=1,2,3,...$) stand for probabilities.
Expanding $\langle A\rangle(\theta_j)$ around $\bar \theta$ as
$$\langle A\rangle (\theta_j)=\langle A\rangle (\bar\theta)+\frac{\partial\langle A\rangle}{\partial\theta}|_{\bar\theta}(\theta_i-\bar\theta)+...$$
and approximating $\langle \Delta A^\dagger\Delta A\rangle$ up to the second order in ($\theta_j-\bar\theta$), we arrive at
\begin{equation}\label{errorP}
\langle \Delta A^\dagger\Delta A\rangle=\frac{\partial\langle A^\dagger\rangle}{\partial\theta}|_{\bar\theta}\frac{\partial\langle A\rangle}{\partial\theta}|_{\bar\theta}(\Delta\theta)^2.
\end{equation}
Here, $(\Delta\theta)^2=\sum_jp_j(\theta_j-\bar\theta)^2$.
This is the error propagation in Eq.(2) of the main text.

With this error propagation, we now derive the non-Hermitian Cram\'er-Rao bound in Eq.(3). \textbf{Introducing an operator $L$, which is not required to be Hermitian,} we have
\begin{eqnarray}\label{crbd1}
(\Delta\theta)^2\frac{\partial\langle A^\dagger\rangle}{\partial\theta} \frac{\partial\langle A\rangle}{\partial \theta} \langle \Delta L^\dagger \Delta L\rangle&=&\langle \Delta A^\dagger \Delta A\rangle\langle \Delta L^\dagger \Delta L\rangle\nonumber\\
&\geq& |\langle A^\dagger L\rangle-\langle A^\dagger\rangle\langle L\rangle|^2.
\end{eqnarray}
\textbf{There are few differences between the present study and the earlier publications in $L$, which deserve to be outlined.} First, in most publications so far, $L$ is the so-called symmetric logarithmic derivative, which is required to be Hermitian. This requirement is due to the fact that the uncertainty relation used in the publications till now is for Hermitian operators.   Whereas in our case $L$ is not required to be Hermitian. In fact, as we show in the main text an anti-Hermitian $L=-L^\dagger$ can maximize the Fisher information. Second, the uncertainty relation used to derive the Cram\'er bound is different, and our derivation can recover the results in the literature.

Until now, the operator $A$ is not required to be Hermitian. As we will show later, a non-Hermitian $A$ will lead a different Cram\'er-Rao bound.

Let us start to discuss a special case that $A$ is Hermitian. For Hermitian $A$,
the last line of Eq. (\ref{crbd1}) becomes,
\begin{eqnarray}\label{ALinequ}
|\langle A^\dagger L\rangle-\langle A^\dagger\rangle\langle L\rangle|^2
&=&|\frac 12 \langle A  L+L^\dagger A\rangle -\langle A \rangle \langle L\rangle+\frac 1 2 \langle A  L-L^\dagger A\rangle|^2\nonumber\\
&\geq& \left (\frac{\langle AL+L^\dagger A\rangle}{2}\right )^2.
\end{eqnarray}
To have the last inequality, we have ignored $\frac 1 2 \langle A  L-L^\dagger A\rangle$, which is pure imaginary, while term $\langle A  L+L^\dagger A\rangle$ is real. By carefully choosing  $L$, we can have $\langle L\rangle=0.$ Eq. (\ref{crbd1}) reduces to Eq.(3) in the main text, where $L$ satisfies Eq.(4).

When $A$ is not Hermitian but with $\langle L\rangle=0$, we still obtain Eq. (3), but in this case, $L$ satisfies  Eq. (5). The proof is almost the same as that for Eq. (4).

Discussions are in order. Either
\begin{equation}\label{LandLd1}
\frac {\partial \rho}{\partial \theta}=\frac 1 2(L^{(1)}\rho+\rho L^{(1)\dagger}),
\end{equation}
or
\begin{equation}\label{LandL2}
\frac {\partial \rho}{\partial \theta}=L^{(2)}\rho=\rho L^{(2)\dagger}.
\end{equation}
guarantee $\langle L\rangle=0.$ For Eq.(\ref{LandLd1}), as we will discuss below, $L^\dagger=L e^{i\beta}$ is assumed. This assumption together with $\Tr \left (\frac{\partial \rho}{\partial \theta}\right )=0$ leads to $\langle L\rangle=0.$ While for Eq.(\ref{LandL2}), it is obvious that
$\langle L\rangle=0$ and $\langle L^\dagger\rangle=0,$ since $\Tr \left (\frac{\partial \rho}{\partial \theta}\right )=0$.

\section{S2.\, non-Hermitian quantum Fisher information}
We present a detailed calculations for the Fisher information $F_{nH}^{(1)}$ and $F_{nH}^{(2)}$ defined in Eq. (3).
We first calculate  $F_{nH}^{(1)}.$  Without lose of generality, we assume the dimension of the Hilbert space is $N$, while the state of the system $\rho$ may be  not of full rank. It has positive eigenvalues $p_j$ required by quantum theory, and the corresponding eigenstates are denoted by $|\phi_j\rangle$, i.e.,
\begin{equation}\label{rhospec}
\rho=\sum_{j=1}^M p_j|\phi_j\rangle\langle \phi_j|.
\end{equation}
Here we assume $j$ runs from $j=1$ to $j=M$, $M\leq N$. With these notations,
the Fishier information defined in Eq.(3) takes,
\begin{eqnarray}\label{fish1}
F_{nH}^{(1)}&=&\langle L^{(1)}L^{(1)} \rangle e^{i\beta}
=\sum_{i=1}^M p_i\langle\phi_i|L^{(1)}L^{(1)}|\phi_i\rangle e^{i\beta}
=\sum_{j=1}^N \sum_{i=1}^Mp_i\langle\phi_i|L^{(1)}|\phi_j\rangle \langle \phi_j |L^{(1)}|\phi_i\rangle e^{i\beta}\nonumber\\
&=&\sum_{j=1}^N \sum_{i=1}^M p_i (L^{(1)})_{ij}(L^{(1)})_{ji}e^{i\beta}
\end{eqnarray}
Recalling Eq. (4) in the main text that,
\begin{equation}\label{LandLd}
\frac {\partial \rho}{\partial \theta}=\frac 1 2(L^{(1)}\rho+\rho L^{(1)\dagger}),
\end{equation}
and assuming  $L^{(1)\dagger}=L^{(1)}e^{i\beta}$, we have
\begin{equation}\label{Lele}
L_{ij}^{(1)}=\langle \phi_i|L^{(1)}|\phi_j\rangle=2\frac{(\partial_\theta\rho)_{ij}}{p_j+p_ie^{i\beta}}.
\end{equation}
As addressed in the main text, $L^{(1)}$ should not be limited to the space spanned by the eigenstates of $\rho$, however for $j>M$ or $i>M$, $L_{ij}^{(1)}$  can not established by  Eq. (\ref{Lele}). We will return to this point later.
To compute $L_{ij}^{(1)}$, we need to know $(\partial_\theta \rho)_{ij}$. Starting from Eq.(\ref{rhospec}), we have
\begin{eqnarray}
\partial_\theta\rho=\sum_i\partial_\theta p_i |\phi_i\rangle\langle\phi_i|+\sum_i p_i|\partial_\theta\phi_i\rangle\langle \phi_i|+\sum_i p_i|\phi_i\rangle\langle \partial_\theta\phi_i|,
\end{eqnarray}
leading to
\begin{eqnarray}\label{rhoij}
(\partial_\theta\rho)_{ij}&=&\partial_\theta p_i \delta_{ij}+ p_i\langle \partial_\theta\phi_i| \phi_j\rangle + p_j\langle \phi_i| \partial_\theta\phi_j\rangle\nonumber\\
&=&\partial_\theta p_i \delta_{ij}+(p_i-p_j)\langle \partial_\theta\phi_i| \phi_j\rangle.
\end{eqnarray}
Here, $\langle \partial_\theta \phi_i|\phi_j\rangle+\langle \phi_i|\partial_\theta\phi_j\rangle=0$ has been applied in the derivation.
Substituting Eq.(\ref{rhoij}) and  Eq. (\ref{Lele}) into Eq. (\ref{fish1}), we have
\begin{eqnarray}
F_{nH}^{(1)}&=&\sum_{i=1}^M \sum_{j=1}^N 4p_i \frac{(\partial_\theta p_i)^2\delta_{ij}}{p_i^2+p_j^2+2p_ip_j \cos\beta}+\sum_{i=1}^M \sum_{j=1}^N 4p_i \frac{(p_i-p_j)^2\langle\partial_\theta\phi_i|\phi_j\rangle \langle \phi_j|\partial_\theta\phi_i\rangle}{p_i^2+p_j^2+2p_ip_j \cos\beta}\nonumber\\
&=&\sum_{i=1}^M \frac{2(\partial_\theta p_i)^2}{p_i(1+\cos\beta)}+\sum_{i=1}^M \sum_{j=1}^M 4p_i \frac{(p_i-p_j)^2\langle\partial_\theta\phi_i|\phi_j\rangle \langle \phi_j|\partial_\theta\phi_i\rangle}{p_i^2+p_j^2+2p_ip_j \cos\beta}+\sum_{i=1}^M \sum_{j=M+1}^N 4p_i \langle\partial_\theta\phi_i|\phi_j\rangle \langle \phi_j|\partial_\theta\phi_i\rangle\nonumber\\
&=&\sum_{i=1}^M \frac{2(\partial_\theta p_i)^2}{p_i(1+\cos\beta)}+\sum_{i=1}^M \sum_{j=1}^M 4p_i \frac{(p_i-p_j)^2\langle\partial_\theta\phi_i|\phi_j\rangle \langle \phi_j|\partial_\theta\phi_i\rangle}{p_i^2+p_j^2+2p_ip_j \cos\beta}\nonumber\\
&+&\sum_{i=1}^M \sum_{j=1}^M 4p_i \langle\partial_\theta\phi_i|\left(\frac {}{}I-|\phi_j\rangle \langle \phi_j|\frac {}{}\right)|\partial_\theta\phi_i\rangle,
\end{eqnarray}
where $I$ is the identity operator of the Hilbert space, i.e., $I=\sum_{i=1}^N|\phi_i\rangle\langle \phi_i|$. Note that the sum runs from 1 to $N$. Finally, we obtain Eq. (8) in the main text,
\begin{eqnarray}
F_{nH}^{(1)}=\sum_{i=1}^M \frac{2(\partial_\theta p_i)^2}{p_i(1+\cos\beta)}+\sum_{i=1}^M 4p_i\langle \partial_\theta\phi_i|\partial_\theta\phi_i\rangle
+\sum_{i=1}^M\sum_{j=1}^M \left( \frac{4p_i(p_i-p_j)^2}{p_i^2+p_j^2+2p_ip_j\cos\beta}-4p_i\right)\langle\partial_\theta\phi_i|\phi_j\rangle
\langle\phi_j|\partial_\theta\phi_i\rangle.\nonumber\\
\end{eqnarray}

$F_{nH}^{(2)}$ can be derived by the same technique as employed in the derivation of $F_{nH}^{(1)}$. Note that
$L^{(2)\dagger}_{ij}=\frac{(\partial_{\theta}\rho)_{ij}}{p_i}$, and $L_{ji}^{(2)}=\frac{(\partial_{\theta}\rho)_{ji}}{p_i}$ for $F_{nH}^{(2)}$.

\section{S3.\, examples }\label{exa}
Detailed calculation for the non-Hermitian quantum Fisher information $F_{nH}^{(1)}$ and $F_{nH}^{(2)}$
will be presented below. For mixed states of form $\rho=\sum_{i=1}^2 p_i|\phi_i\rangle\langle\phi_i|$ with $\theta$-independent $p_i$,
the Fisher information $F_H$, $F_{nH}^{(1)}$ and $F_{nH}^{(2)}$ reduce to,
\begin{eqnarray}\label{Fmixed}
F_H&=&\sum_{i=1}^2 \left(4p_i\langle\partial_\theta\phi_i|\partial_\theta\phi_i\rangle
-4p_i|\langle\phi_i|\partial_\theta\phi_i\rangle|^2\right)
-16p_1p_2|\langle\phi_1|\partial_\theta\phi_2\rangle|^2,\nonumber\\
F_{nH}^{(1)}&=&\sum_{i=1}^2 4p_i\langle\partial_\theta\phi_i|\partial_\theta\phi_i\rangle,\, (\beta=\pi),\nonumber\\
F_{nH}^{(2)}&=&\sum_{i=1}^2 \left(p_i\langle\partial_\theta\phi_i|\partial_\theta\phi_i\rangle
-p_i|\langle\phi_i|\partial_\theta\phi_i\rangle|^2\right)
+\left(\frac{p_2(1-3p_1)}{p_1}+\frac{p_1(1-3p_2)}{p_2}\right)
|\langle\phi_1|\partial_\theta\phi_2\rangle|^2.
\end{eqnarray}

In the first example, the state before encoding is
\begin{eqnarray}
\rho&=&\sum_{i=1}^2 p_i|\phi_i\rangle\langle \phi_i|,\nonumber\\
|\phi_{1,2}\rangle&=&\frac {1} {\sqrt 2} (|a\rangle^{\otimes N}\pm |b\rangle^{\otimes N}).
\end{eqnarray}
The signal Hamiltonian used to encode the parameter into the state is $J_z$, i.e., the encoding operation is  given by $U(\theta)=e^{-i\theta  J_z}$. The states that carry the information of the parameter is then
\begin{eqnarray}\label{encostate1}
\rho(\theta)&=&\sum_{i=1}^2 p_i|\phi_i(\theta)\rangle\langle \phi_i(\theta)|,\nonumber\\
|\phi_{1,2}(\theta)\rangle&=&\frac {1} {\sqrt 2} (e^{\frac i 2 N\theta}|a\rangle^{\otimes N}\pm e^{-\frac i 2 N\theta}|b\rangle^{\otimes N}).
\end{eqnarray}
It is easy  to find that
\begin{eqnarray}
\langle\partial_\theta\phi_1|\partial_\theta\phi_1\rangle&=&\frac 1 4 N^2,\nonumber\\
\langle\partial_\theta\phi_2|\partial_\theta\phi_2\rangle&=&\frac 1 4 N^2,\nonumber\\
\langle\partial_\theta\phi_1|\phi_1\rangle=\langle\partial_\theta\phi_2|\phi_2\rangle&=&0,\nonumber\\
|\langle\partial_\theta\phi_1|\phi_2\rangle|^2&=&\frac 1 4 N^2.
\end{eqnarray}
Finally, we obtain Eq.(12) of the main text.

\begin{figure}
\includegraphics*[width=0.8\columnwidth,height=0.6\columnwidth]{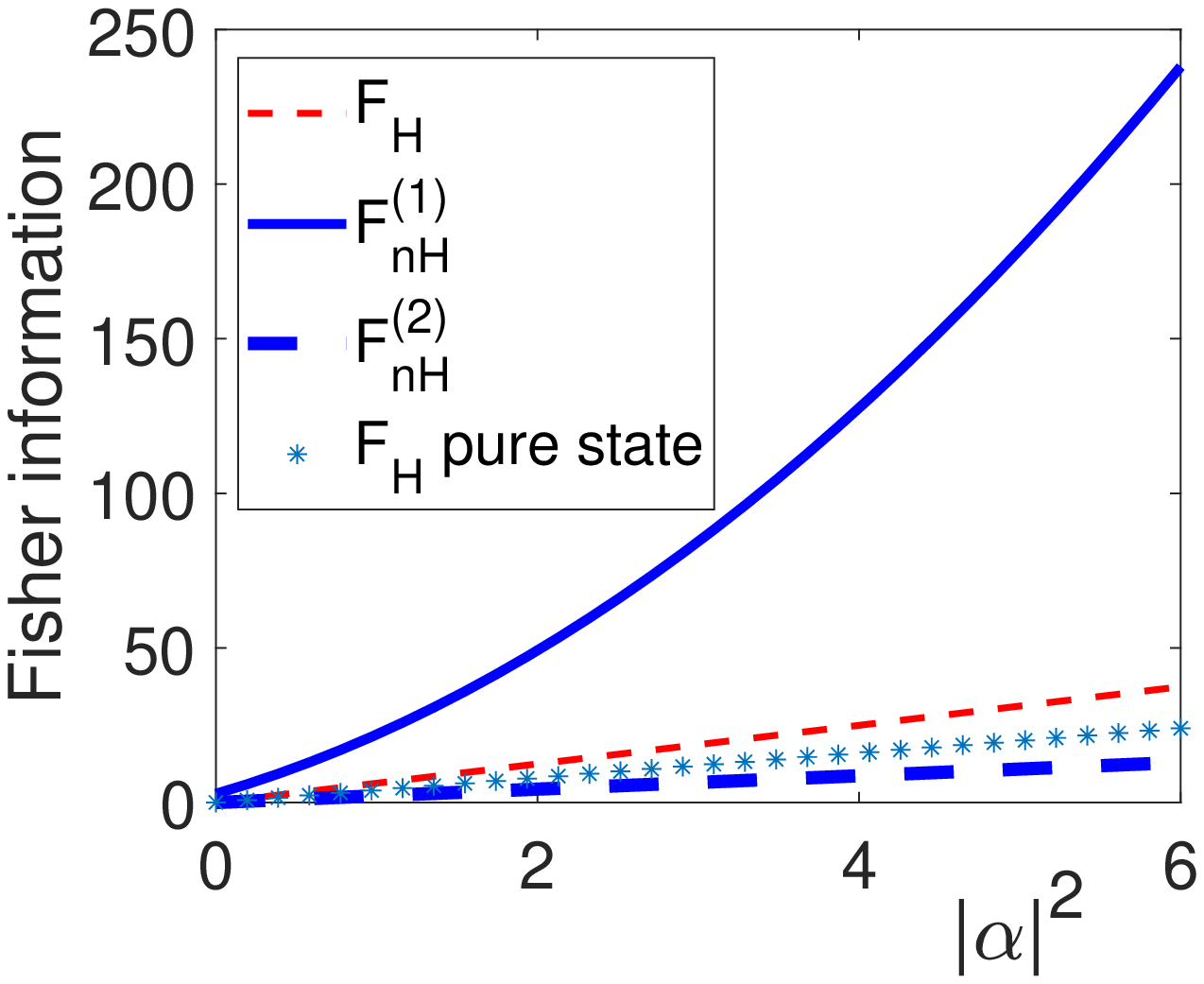} \caption{Fisher information as a function of photon number. $p=0.3$. This is the numerical simulation for Eq.(14) in the main text.} \label{qcbsca2}
\end{figure}

In the second example, the state before encoding is
\begin{eqnarray}
\rho&=&\sum_{i=1}^2 p_i|\phi_i\rangle\langle \phi_i|,\nonumber\\
|\phi_{1}\rangle&=& |r\alpha\rangle_{00}\equiv S(r)\otimes D(\alpha)|00\rangle,\nonumber\\
|\phi_{2}\rangle&=& |r\alpha\rangle_{01}\equiv S(r)\otimes D(\alpha)b^\dagger|00\rangle,
\end{eqnarray}
where $|00\rangle$ denotes that both modes $a$ and $b$ are in its   vacuum state. $S(r)$ and $D(\alpha)$ was defined in the main text, i.e., $S(r)=\exp(\frac {r^*}{2}a^2-\frac{r}{2}(a^{\dagger })^2)$, $ D(\alpha)=\exp(\alpha a^\dagger-\alpha^* a)$, and we will set $r=|r|e^{i2\phi}$. The parameter is encoded into the state after passing through a beam splitter represented by $B_\pi$. The state encoding the parameter $\theta$ is then
\begin{eqnarray}
\rho(\theta)&=&\sum_{i=1}^2 p_i|\phi_i(\theta)\rangle\langle \phi_i(\theta)|,\nonumber\\
|\phi_{1,2}(\theta)\rangle&=&U(\theta)BS(r)D(\alpha)|0\rangle_a\otimes |X_{1,2}\rangle_b,
\end{eqnarray}
where $|X_1\rangle_b=|0\rangle_b$, $|X_2\rangle_b=b^{\dagger}|0\rangle_b,$ and $p_1+p_2=1.$
Further, we find
\begin{eqnarray}
\langle\partial_\theta\phi_1|\partial_\theta\phi_1\rangle&=&|\alpha|^4+|\alpha|^2,\nonumber\\
\langle\partial_\theta\phi_2|\partial_\theta\phi_2\rangle&=&|\alpha|^4+5|\alpha|^2+1,\nonumber\\
|\langle\partial_\theta\phi_1|\phi_1\rangle|^2&=&|\alpha|^4,\nonumber\\
|\langle\partial_\theta\phi_2|\phi_2\rangle|^2&=&(1+|\alpha|^2)^2,\nonumber\\
|\langle\partial_\theta\phi_1|\phi_2\rangle|^2&=&|\alpha|^2.
\end{eqnarray}
Collecting these results, we arrive at
\begin{eqnarray}\label{Sexa1}
F_H&=&4|\alpha|^2(4p^2-6p+3), \nonumber\\
F_{nH}^{(1)}&=&4|\alpha|^4+(20-16p)|\alpha|^2+4(1-p),\nonumber\\
F_{nH}^{(2)}&=&\frac{2p^3-2p^2+1}{p(1-p)}|\alpha|^2.
\end{eqnarray}
This is Eq.(14) in the main text. The dependence of the Fisher information on the photon number is depicted in Fig.\ref{qcbsca2}.
The Hermitian Fisher information for pure state is also shown for comparison with the others. We can clearly find that for large photon number, $F_{nH}^{(1)}$ scales dominantly with $|\alpha|^4=N^2$, indicating the breakdown of the Heisenberg limit.

\section{S4.\, non-Hermitian signal Hamiltonian}
In the following, we will consider a situation that the signal Hamiltonian is not Hermitian for $F_{H}$. Comparison is carried out between the Hermitian quantum Fisher information $F_H$ with Hermitian signal Hamiltonian and the Hermitian quantum Fisher information with non-Hermitian signal  Hamiltonian (will be denoted by $F_{H}^{nhs}$), and only the case of pure states is considered.

For pure state $|\phi\rangle=|\phi(\theta)\rangle$, $F_{nH}^{(1)}=F_{H}$, so we do not take  $F_{nH}^{(1)}$ into comparison. For pure states,  $F_{H}^{nhs}$ and $F_H$ takes,
\begin{eqnarray}
F_H&=& 4\langle \partial_\theta\phi|\partial_\theta\phi\rangle
-4\langle\partial_\theta\phi|\phi\rangle
\langle\phi|\partial_\theta\phi\rangle,\label{nnF}\\
F_{H}^{nhs}&=&4\left(M^2\langle\partial_\theta\phi^\prime|\partial_\theta \phi^\prime\rangle-\frac {1}{M^2}(\partial_\theta M)^2-\left|\frac{\partial_\theta M}{M}+M^2\langle\phi^\prime|\partial_\theta\phi^\prime\rangle\right|^2\right),\label{nnF2}
\end{eqnarray}
where $|\phi^\prime\rangle$ denotes the unnormalized encoding state and $M$ is its normalization constant, namely, $\langle\phi^\prime|\phi^\prime\rangle=\frac {1}{M^2}$. We might prove Eq.(\ref{nnF2}) by substitution of  $|\phi\rangle=M|\phi^\prime\rangle$ into  $F_{H}=4\langle \partial_\theta\phi|\partial_\theta\phi\rangle
-4\langle\partial_\theta\phi|\phi\rangle
\langle\phi|\partial_\theta\phi\rangle$,  and noticing
\begin{eqnarray}
\langle\partial_\theta\phi|\partial_\theta\phi\rangle&=&M^2\langle\partial_\theta\phi^\prime
|\partial_\theta\phi^\prime\rangle-\frac{(\partial_\theta M)^2}{M^2},\nonumber\\
\langle\phi|\partial_\theta\phi\rangle&=&\frac{\partial_\theta M}{M}+M^2\langle\phi^\prime|\partial_\theta\phi^\prime\rangle.
\end{eqnarray}
In the first example, the signal Hamiltonian $J_z$ is replaced with $J_z(1-i\gamma)$, the encoding state would have the same form as in Eq.(\ref{encostate1}) but with $\bar N=N(1-i\gamma)$ instead of $N$. With these knowledge, we have
\begin{eqnarray}
F_H&=&N^2,\nonumber\\
F_{H}^{nhs}&=&{|\bar N|^2}(1-\tanh^2(\gamma N\theta)),
\end{eqnarray}
which is Eq.(15) in main text.
\begin{figure}
\includegraphics*[width=0.8\columnwidth,height=0.6\columnwidth]{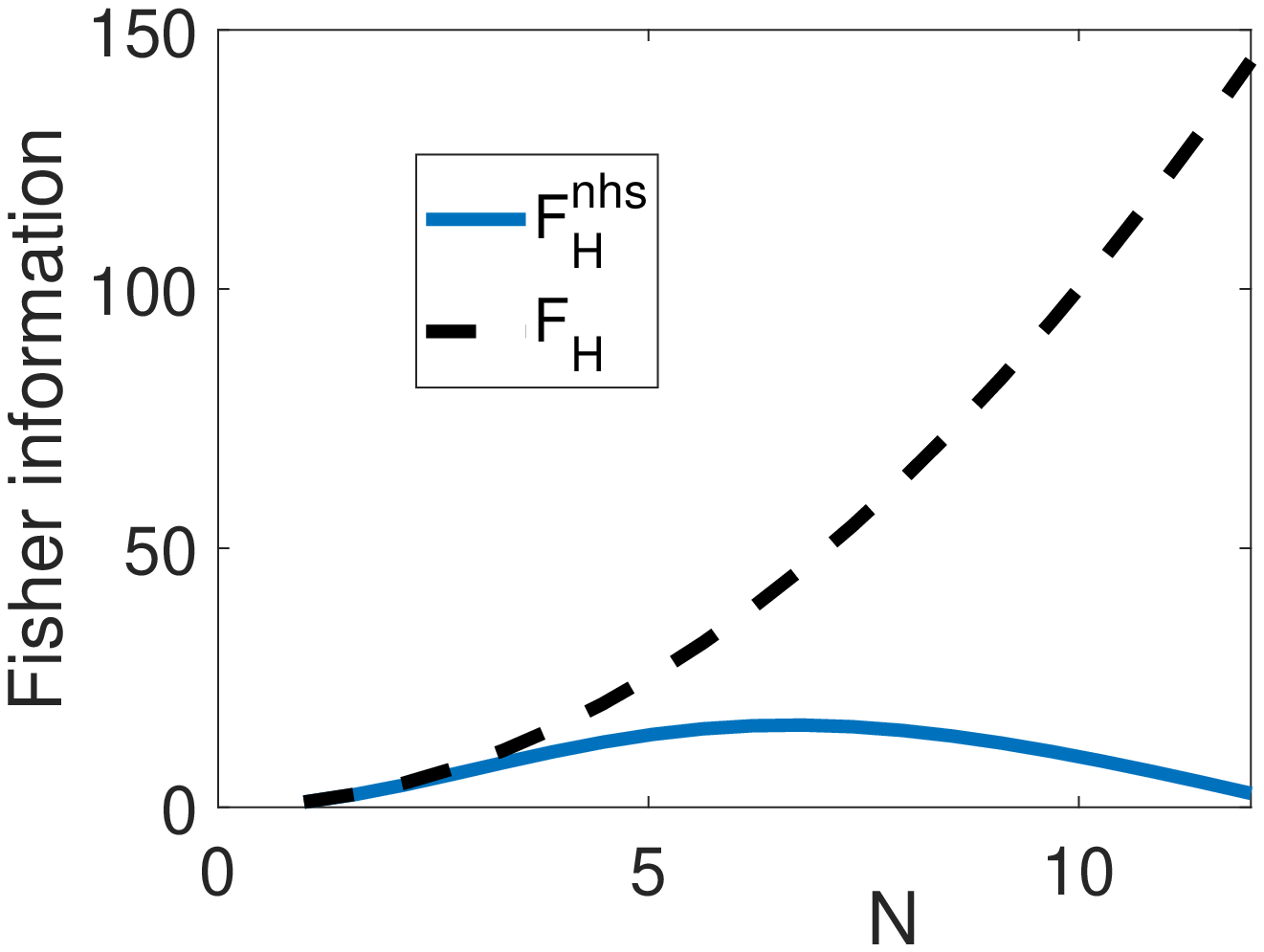} \caption{Fisher information as a function of ion number.  This is the numerical simulation for Eq.(15) in the main text. The parameters chosen are $\theta=0.25\pi$ and
$\gamma=0.2$.} \label{qcbsca1}
\end{figure}

One might wonder why the signal Hamiltonian takes $J_z(1-i\gamma)$ instead of $J_z$. This can be understood as follows. Consider the rotation $U(\theta)=e^{-i\theta  J_z}$ applied to encode the parameter $\theta$ into the state. This rotation can be realized in trapped ions\cite{monz11} through an effective Hamiltonian $H_{signal}$,
\begin{equation}
H_{signal}=\frac{\hbar \Omega}{2}J_z,
\end{equation}
where $J_z=\sum_l s_z^{(l)}$ is a collective spin of those trapped ions, $\Omega$ is the Rabi frequency of each ion or the magnetic field $B_z$ to which the ion coupled.
The time evolution operator $U(T)=e^{-\frac{i}{\hbar}H_{singal}T}\equiv e^{-i\theta  J_z}$ plays the role of the rotation operator.
Here the accumulated phase $\theta=\frac{\Omega}{2}T.$

There is no quantum system being completely isolated from its surroundings including the trapped ions. A system that is coupled with its environment is called open system.
A complete description of an open quantum system requires the inclusion of the environment. As a result of the interaction with the environment, the dynamics of the open system is governed by the master equation\cite{gardiner04},
\begin{eqnarray}
\frac{\partial\rho}{\partial t}&=&-\frac i \hbar[H_{signal},\rho]+\cal{L}(\rho),\nonumber\\
\cal{L}(\rho)&=&\frac\kappa 2\sum_l(2s_-^{(l)}\rho s_+^{(l)}-\rho s_+^{(l)}s_-^{(l)}-s_+^{(l)}s_-^{(l)}\rho).
\end{eqnarray}
Consider a very short encoding time, terms with $s^{(l)}_-\rho s^{(l)}_+$ can be ignored.
And the dynamics of the open system is governed by an effective Hamiltonian
\begin{equation}
H_{eff}=H_{signal}-\frac{i\hbar\kappa}{2} \sum_ls^{(l)}_+s^{(l)}_-.
\end{equation}
Consider  each ion being modelled by a two-level system, the terms $\sum_ls^{(l)}_+s^{(l)}_-$ would contribute to $H_{signal}$ as $\sim J_z$. Namely, in this open system
\begin{equation}
\bar H_{singal}\simeq J_z(1-i\gamma).
\end{equation}
Here $\gamma\sim \kappa$ stands for the decay rate of collective spin  $J_z$.

For the example with Mach-Zehnder interferometer, we take $\bar H=a^{\dagger}a(1-i\gamma)$  to replace $H$ in the Hermitian encoding case. The calculation is involved but straightforward, it shows that
\begin{eqnarray}\label{ex2nnh}
M^2&=&\exp\left[\frac{}{}|\alpha|^2(1-e^{-2\gamma\theta})\right],\nonumber\\
\langle\partial_\theta \phi^\prime|\partial_\theta \phi^\prime\rangle&=&\frac{1+\gamma^2}{M^2}|\alpha|^2
e^{-2\gamma\theta}(1+|\alpha|^2e^{-2\gamma\theta}),\nonumber\\
\langle\partial_\theta \phi^\prime | \phi^\prime\rangle&=&(i-\gamma)\frac{|\alpha|^2}{M^2}e^{-2\gamma\theta}.
\end{eqnarray}
Substituting Eq.(\ref{ex2nnh}) into Eq.(\ref{nnF2}), we obtain Eq.(16) in the main text.

\section{S5.\, results with the other beam splitter}

\begin{figure}
\includegraphics*[width=0.9\columnwidth,height=0.6\columnwidth]{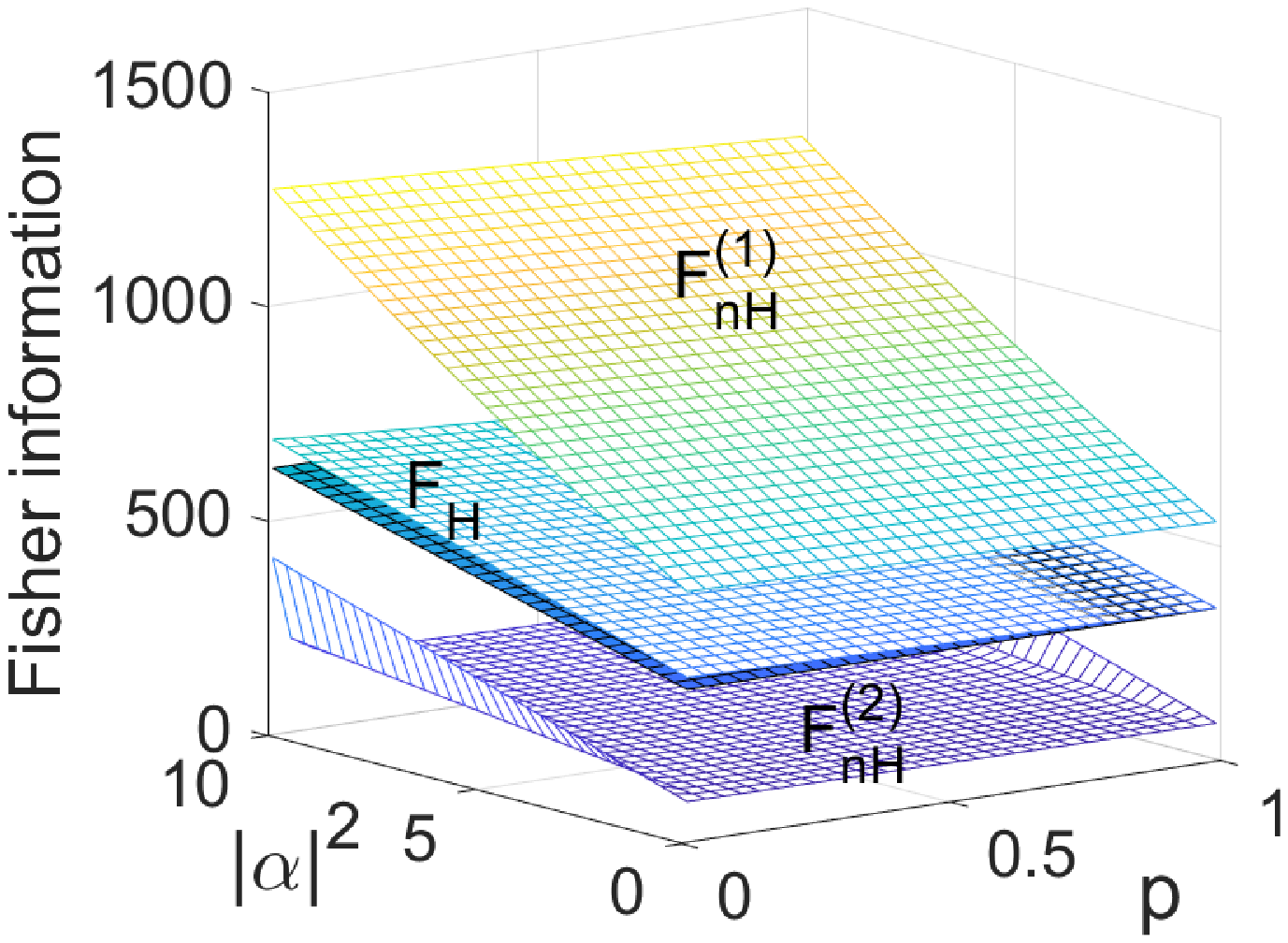} \caption{Fisher information as a function of $|\alpha|^2$.  This is the numerical simulation for Fisher information with beam splitter $B_{\frac\pi 2}$. The parameters chosen are $\phi=0.25\pi$ and
$|r|=2$. The surf figure just below $F_H$ (mesh figure) is for the Hermitian quantum information in pure state $|\phi_1\rangle$.} \label{qcbsf1_B}
\end{figure}

In Sec.\ref{exa},  we employed a beam splitter, $B_\pi=e^{-\frac i 2 \pi (ab^\dagger+a^\dagger b)}$,
in a standard Mach-Zehnder setup to study the Fisher information. With beam splitter $B_\pi$, the mode $b$ (or mode $a$) is simply transmitted to another mode $a$ (or mode $b$), so there is no interferometer at all, the Fisher information is however not zero \cite{jarzyna12}. In the following,
we will employ the other beam splitter,
$B_{\frac \pi 2}=e^{-\frac i 4 \pi (ab^\dagger+a^\dagger b)}$, to study the Fisher information.
Involved but straightforward calculations show that,
\begin{eqnarray}\label{B2mz}
\langle\partial_\theta \phi_2|\partial_\theta\phi_2\rangle
&=&\langle\partial_\theta \phi_1|\partial_\theta\phi_1\rangle+\sinh^2 |r|+|\alpha|^2+\frac 1 2,\nonumber\\
\langle\partial_\theta \phi_1|\partial_\theta\phi_1\rangle&=&
\frac 1 4\left(|\alpha|^4+2|\alpha|^2+4|\alpha|^2\sinh^2|r|+\sinh^4|r|+2\sinh^2|r|\cosh^2|r|+
\sinh^2|r|\right)\nonumber\\
&+&\frac 1 2\Re(\alpha^2e^{-2i\phi})\cosh |r|\sinh |r|,\nonumber\\
|\langle\phi_1|\partial_\theta\phi_1\rangle|^2&=&\frac 1 4 (\sinh^2|r|+|\alpha|^2)^2,\nonumber\\
|\langle\phi_1|\partial_\theta\phi_2\rangle|^2&=&\frac 1 4 |\alpha|^2,\nonumber\\
|\langle\phi_2|\partial_\theta\phi_2\rangle|^2&=&\frac 1 4 (\sinh^2|r|+|\alpha|^2+1)^2.
\end{eqnarray}
Here $\Re(...)$ stands for the real part of $(...)$. Substituting Eq.(\ref{B2mz}) into Eqs(\ref{Fmixed}), we obtain the Fisher information of states with beam splitter $B_{\frac \pi 2}$. Numerical results are presented in Fig. \ref{qcbsf1_B}. We find that $F_{nH}^{(2)}<F_H<F_{nH}^{(1)}$ except $p\rightarrow 0, 1$. This is  similar to the results in Fig. 2 of the main text.

\section{S6.\, Saturation of non-Hermitian quantum Cram\'er-Rao bounds $1/F_{nH}^{(1)}$ and $1/F_{nH}^{(2)}$}
The variance  $(\Delta\theta)^2$ of the estimated parameter $\theta$ is given by the error propagation Eq.(2) of the main text, it is  bounded by the quantum
Cram\'er-Rao bound (QCRB) defined through the quantum Fisher information as
$$(\Delta\theta)^2\geq \frac{1}{F}, \quad F=F_H, \, F_{nH}^{(1)}, \, F_{nH}^{(2)}.$$
Given a signal Hamiltonian and initial state, the bounds can be saturated by carefully chosen   measurements characterized by variable (operator) $A$. In the following, we will show that the bound given by $F_{nH}^{(2)}$ can be saturated, while $F_{nH}^{(1)}$ can not. The optimal measurement $A^{opt}$ to saturate the bound $\frac{1}{F_{nH}^{(2)}}$ is also given.

Recalling the Cauchy-Schwarz inequality, $\langle F|F\rangle \langle G|G\rangle \geq |\langle F|G\rangle |^2$ and noticing the equality holds if and only if $|F\rangle = C |G\rangle,$ ($C$ is a constant) we claim that
to saturate the bounds, $A^{opt}=\Gamma L$ ($\Gamma$ is a constant). In other words, to reach the lower Cram\'er-Rao  bound, the optimal measurement variable $A^{opt}$ is required to be proportional to the symmetric logarithmic derivative $L$.

In Eq. (\ref{ALinequ}), a term $\langle A^\dagger L-L^\dagger A\rangle$ had been ignored in order to have the last inequality. For the equality to hold, $(A^{opt})^{\dagger} L=L^\dagger A^{opt}$ is required, this together with $A^{opt}=\Gamma L$ require that $\Gamma$ is real. This means that to saturate the  bounds, $A^{opt}$ and $L$ must be
Hermitian or non-Hermitian simultaneously. This suggests that bound $1/F_{nH}^{(1)}$ can not be saturated, as $A$ is Hermitian but $L$ is not as we stated  in the main text. In addition, $\beta=\pi$ can not be met as discussed in the main text.

As to the bound given by $1/F_{nH}^{(2)}$, all requirements for  saturation are met, and the optimal measurement $A^{opt}$ can be expressed  in terms of the encoding state $\rho(\theta)$. Firstly, $(A^{opt})^{\dagger} L=L^\dagger A^{opt}$ and $A^{opt}=\Gamma L$ can be met simultaneously, and there are no contradictions with the requirement in the main text.  Secondly, from $A^{opt}=\Gamma L$ and Eq. (\ref{LandL2}), we find that
\begin{eqnarray}\label{Aopt}
A^{opt}_{ij}&\equiv& \langle \phi_i|A^{opt}|\phi_j\rangle=\Gamma L_{ij},\nonumber\\
L_{ij}&=&\frac{(\partial_{\theta}\rho)_{ij}}{p_j},
\end{eqnarray}
where, $p_i$ and $|\phi_i\rangle$ are the eigenvalues and its corresponding eigenstates of the encoding state $\rho(\theta)$, respectively.

\begin{figure}
\includegraphics*[width=0.9\columnwidth,height=0.6\columnwidth]{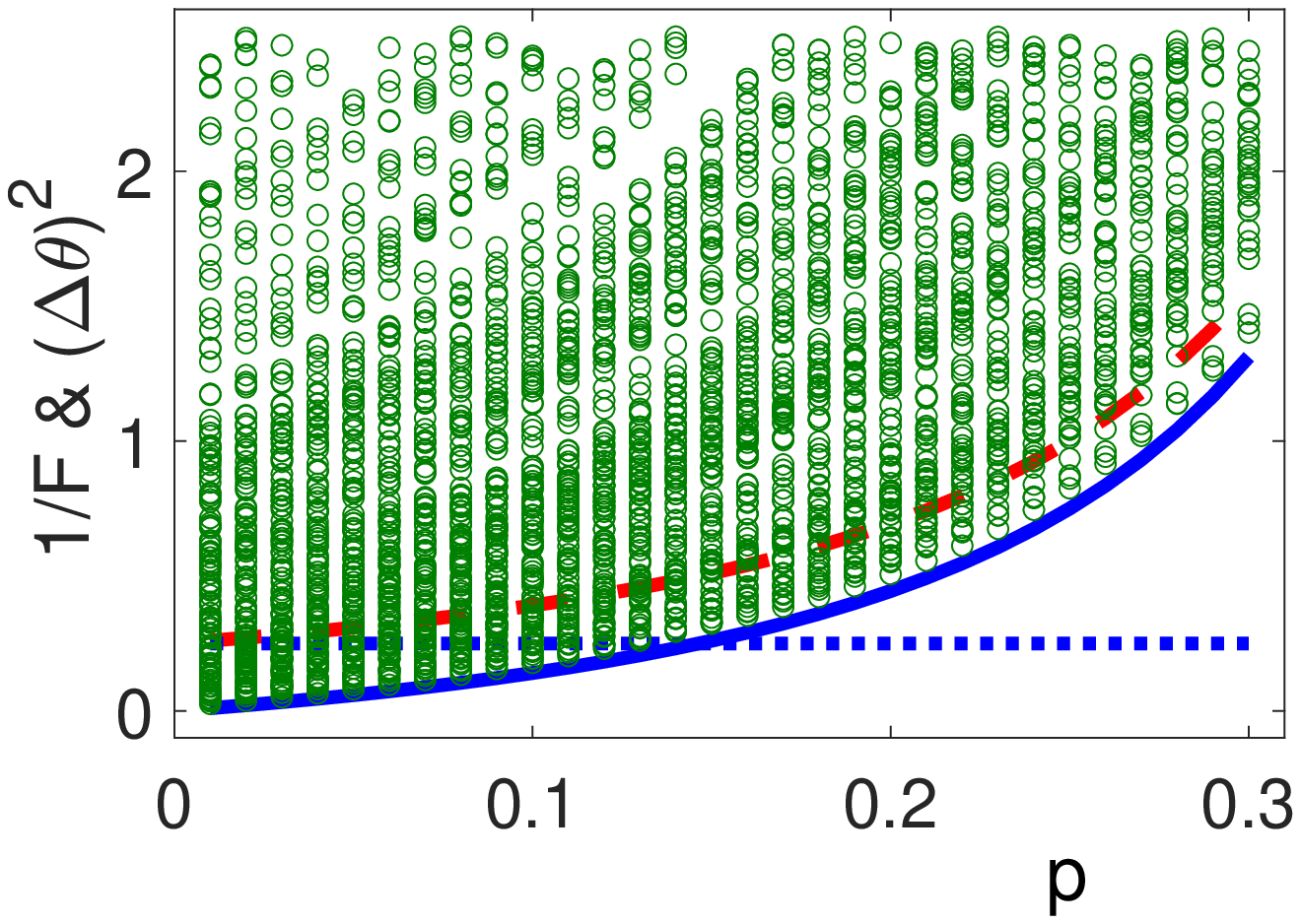} \caption{Bounds $1/F_H$ (red-dashed), $1/F_{nH}^{(1)}$ (blue-dotted) and  $1/F_{nH}^{(2)}$ (blue-solid) as well as the variance of the estimated parameter $(\Delta\theta)^2$ (green-circles) versus $p$. $p$ is defined in Eq. (\ref{Sexa1}), by which we plot the bounds. The variance (green-circles) is calculated by randomly generating variable $A$ and computing $(\Delta\theta)^2$ with Eq. (\ref{errorP}).} \label{boundsAv}
\end{figure}

\begin{figure}
\includegraphics*[width=0.9\columnwidth,height=0.6\columnwidth]{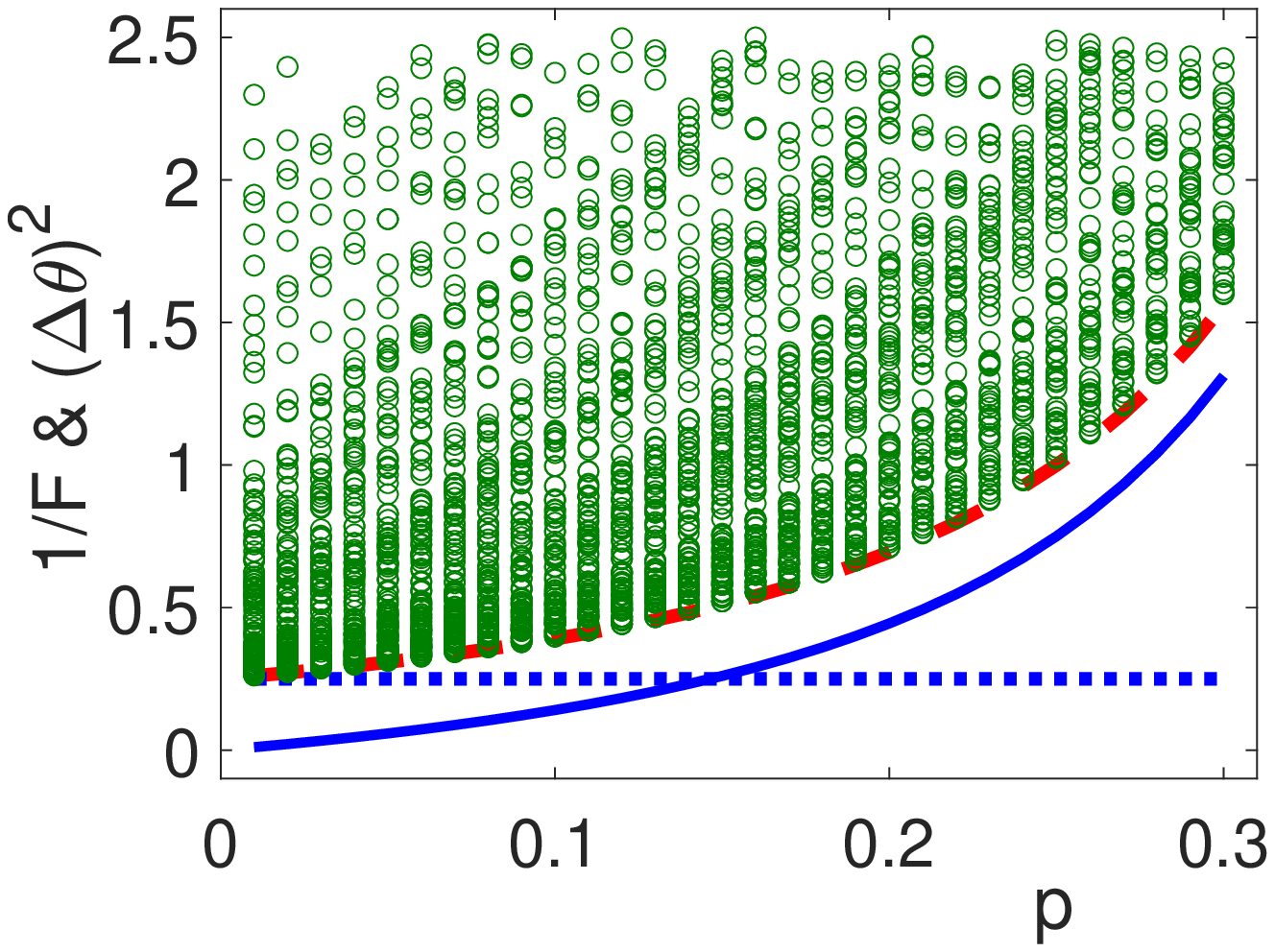} \caption{The same as Fig.
\ref{boundsAv}. The difference is that the optimal $A^{opt}$ here is for the saturation of $1/F_H$.  } \label{boundsAv2}
\end{figure}

To be specific, we assume $p_i$ to be $\theta$-independent as we did in the main text, then
\begin{eqnarray}
(\partial_\theta \rho)_{ij}=p_i\langle\partial_\theta\phi_i|\phi_j\rangle+p_j\langle\phi_j|\partial_\theta\phi_i\rangle.
\end{eqnarray}
We apply these results to the first example, and find that the operators given in Eq. (\ref{Aopt}) indeed saturate the bound $1/F_{nH}^{(2)}$, see Fig. \ref{boundsAv}. The saturation of bound $1/F_H$ is also presented for comparison, see Fig.  \ref{boundsAv2}.

\begin{figure}
\includegraphics*[width=0.9\columnwidth,height=0.6\columnwidth]{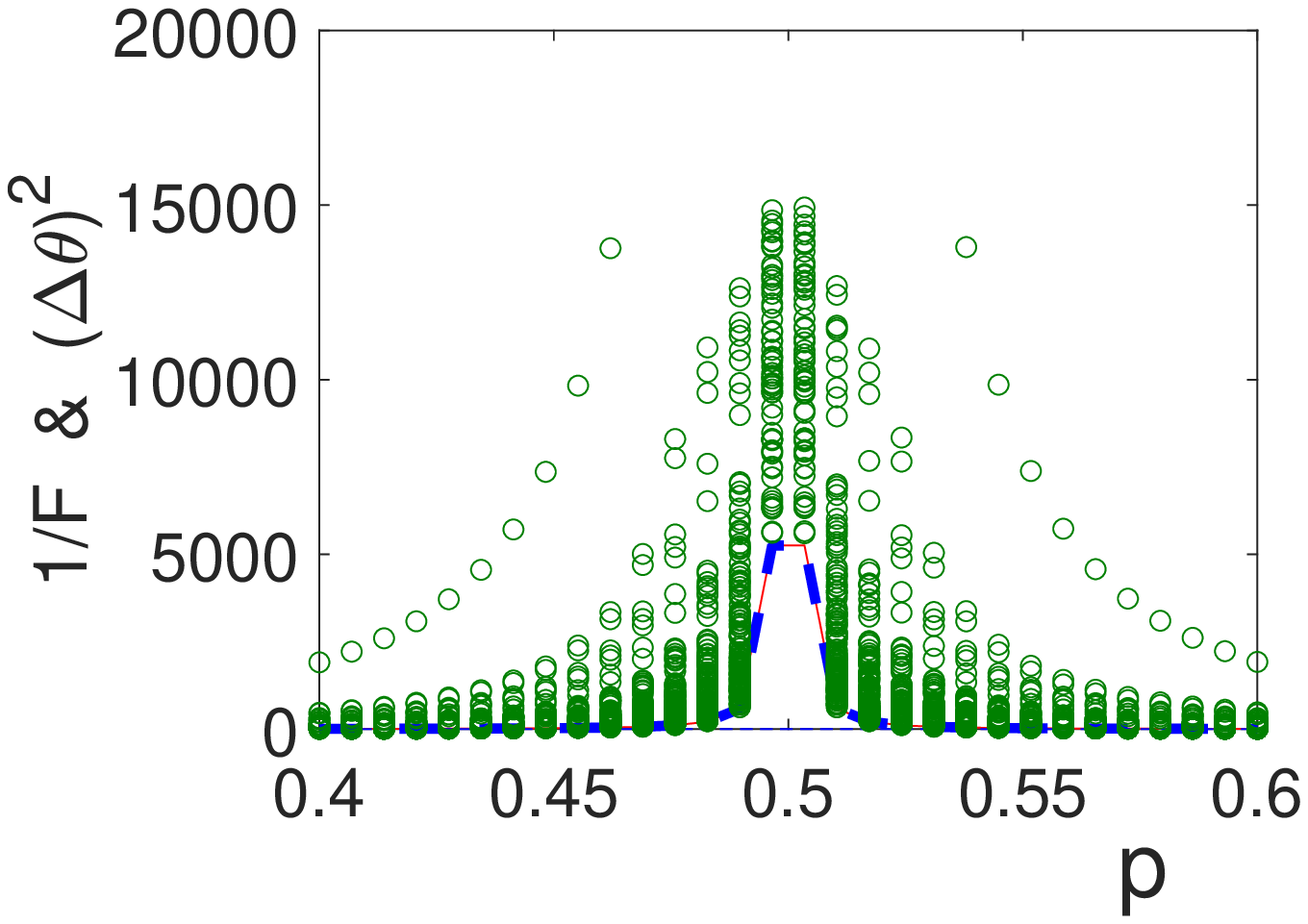} \caption{The same as Fig.
\ref{boundsAv}. The bounds $1/F_H$ (red-solid) and $1/F_{nH}^{(2)}$ (blue-dashed) coincide, while $1/F_{nH}^{(1)}$ is very small with respect to $1/F_H$   and $1/F_{nH}^{(2)}$. } \label{boundsAv3}
\end{figure}

We would like to notice  that for $0.4<p<0.6$, the bounds $1/F_{nH}^{(2)}$ and $1/F_{H}$ are almost  same, while $1/F_{nH}^{(1)}$ remains unchanged. See Fig. \ref{boundsAv3}. For $p>0.6$ the bounds and the variance behave similarly with respect the those in the region  of $p<0.4$.

\begin{figure}
\includegraphics*[width=0.9\columnwidth,height=0.6\columnwidth]{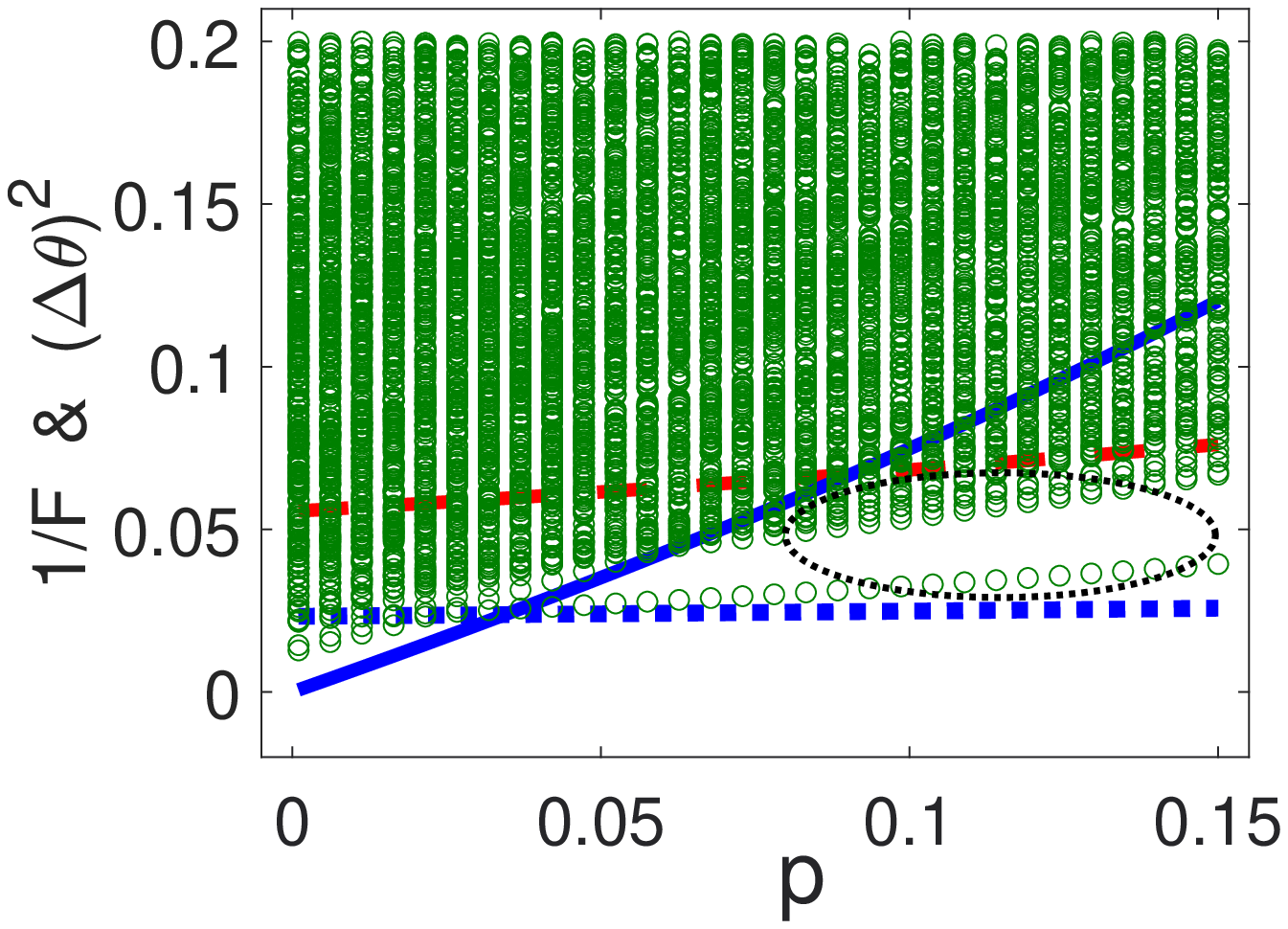} \caption{The bounds $1/F_H$ (red-dashed), $1/F_{nH}^{(2)}$ (blue-solid) and $1/F_{nH}^{(1)}$ (blue-dotted) as well as the variance $(\Delta\theta)^2$ (green-circles)  as a function of $p$.  This figure shows the relation between the bounds and the variance in the second example, i.e., the Mach-Zehnder setup. We can find that bound $1/F_{nH}^{(1)}$ can not be saturated, however it really provides a lower bound for the variance $(\Delta \theta)^2$.} \label{boundsA_bs_nH}
\end{figure}

From Figures \ref{boundsAv}, \ref{boundsAv2} and \ref{boundsAv3} we find that the bound $1/F_{nH}^{(1)}$ plays no role yet in  constraint on the parameter estimation. To show the
role that $1/F_{nH}^{(1)}$ plays, we examine the  Mach-Zehnder setup with $B_\pi$ splitter. The results are presented in Fig. \ref{boundsA_bs_nH}. The circles in the region enclosed by the red-dashed, blue-solid and the blue-dashed lines (i.e., the region  with dotted ellipse) break the constraint by $1/F_{H}$ and $1/F_{nH}^{(2)}$ while they are above the lower bound given by $1/F_{nH}^{(1)}$.

\section{S7.\, Measuring the average of a non-Hermitian operator}
The key point of parameter estimation with non-Hermitian systems is to measure the average of non-Hermitian operator $A$. In this section, we follow the proposal in Refs \cite{pati15,nirala19,abbott19,sahoo20} to show how to measure the non-Hermitian operator in our case.

Let us consider a non-Hermitian operator $A$. The expectation
value of $A$ in a quantum state $|\phi_{in}\rangle$ given
by $\langle\phi_{in}|A|\phi_{in}\rangle$ is  complex  in general. This makes the non-Hermitian operator $A$
unobservable in experiments. Nevertheless, recent studies shown that this obstacle can be overcome
with the help of polar decomposition \cite{hall15}, which states that  given any operator $A$, we can always  decompose $A$ as $A = UR$ with  $U$ being an  unitary operator and $R$  a
Hermitian semidefinite operator, $R=\sqrt{A^\dagger A}.$ This bridges the average of non-Hermitian operator $A$ and the weak value of Hermitian operator $R$ as follows,
\begin{eqnarray}
\langle \phi_{in} |A|\phi_{in}\rangle=\langle \phi_{in} |UR|\phi_{in}\rangle=
\frac{\langle \phi |R|\phi_{in}\rangle}{\langle \phi |\phi_{in}\rangle}\langle \phi |\phi_{in}\rangle,
\end{eqnarray}
where $\langle \phi |\equiv \langle \phi_{in} |U.$  It is Well-known that $\frac{\langle \phi |R|\phi_{in}\rangle}{\langle \phi |\phi_{in}\rangle}$ is a weak value of the
positive-semidefinite operator $R$, which can be measured directly in the weak measurement \cite{aharonov88}.

In fact, the weak measurement is not necessary for the weak value and the average of non-Hermitian operator $A$---they can be given in a  single-shot measurement with an interferometric technique reported in \cite{sahoo20}. Before going into details of such a technique, we first show how to decompose the non-Hermitian operator $A$ in our first model.

To simplify the representation, let us consider a single trapped ion. For such a system, any non-Hermitian operator $A$ can be written as,
\begin{equation}
A=a\sigma_++b\sigma_-+c\sigma_z+d,
\end{equation}
where $a, b, c, d$ are complex parameters in general, and $\sigma_+=|a\rangle\langle b|=(\sigma_-)^\dagger,$ $\sigma_z=|a\rangle\langle a|-|b\rangle\langle b|.$ It is easy to find that
\begin{equation}
A^\dagger A=A_x\sigma_x+A_y\sigma_y+A_z\sigma_z+A_0,
\end{equation}
where $A_j,\, j=x,y,z,0$ can be written in terms of $a, b, c, d$ and are all real since $A^\dagger A$ is Hermitian. Namely,   $A_x=\Re(b^*d-b^*c+ac^*+ad^*),$ $A_y=-\Im(b^*d-b^*c+ac^*+ad^*),$ $A_z=2{\cal R}(c^*d)+\frac 1 2 (|b|^2-|a|^2),$ $A_0=\frac 1 2 (|a|^2+|b|^2)+|c|^2+|d|^2.$ $\Re(...)$ and $\Im(...)$ denote the real and imaginary part of $(...).$
Simple algebra yields the eigenstates and the corresponding eigenvalues of $A^\dagger A$,
\begin{eqnarray}
|E^A_{+}\rangle&=&\cos\frac{\Theta}{2}e^{-i\Xi}|a\rangle+ \sin\frac{\Theta}{2}|b\rangle,\nonumber\\
|E^A_{-}\rangle&=&\sin\frac{\Theta}{2}e^{-i\Xi}|a\rangle- \cos\frac{\Theta}{2}|b\rangle,\nonumber\\
E^A_{\pm}&=&A_0\pm\sqrt{A_x^2+A_y^2+A_z^2}.
\end{eqnarray}
Here, $\tan\Xi=\frac{A_y}{A_x},$ and $\cos\Theta=\frac{A_z}{\sqrt{A_x^2+A_y^2+A_z^2}}.$
Then we have, $A^\dagger A=\sum_{j=+,-} E^A_{j}|E^A_j\rangle\langle E^A_j|,$ leading to the decomposition,
\begin{equation}
R=\sqrt{A^\dagger A}=\sum_{j=+,-} \sqrt{E^A_{j}}|E^A_j\rangle\langle E^A_j|.
\label{rr}
\end{equation}
Note that $E^A_{\pm}\geq 0$ as $A^\dagger A$  is Hermitian and  positive. This observation together with $A=UR$ leads to
\begin{equation}
U=A \sum_{j=+,-}\frac{1}{\sqrt{E^A_{j}}}|E^A_j\rangle\langle E^A_j|.
\label{uu}
\end{equation}

\begin{figure}
\includegraphics*[width=0.9\columnwidth,height=0.6\columnwidth]{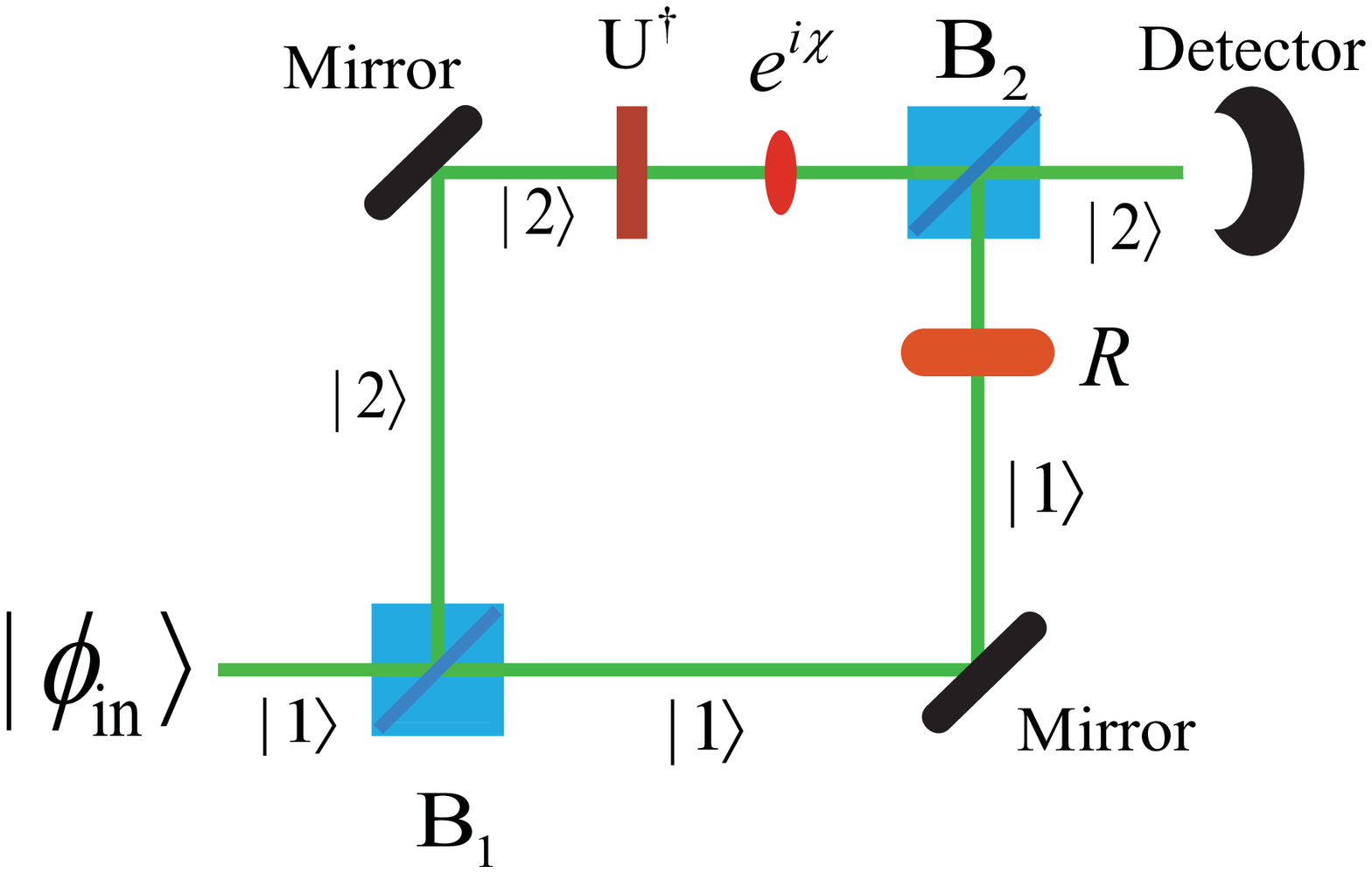} \caption{Schematic setup for measurement of non-Hermitian operator $A=UR.$ $B_1$ and $B_2$ are 50:50 Hadamard-type beam splitters, which split the spatial modes representing by $|1\rangle$ and $|2\rangle$. $e^{i\chi}$ is a phase shifter that  introduces a relative phase $\chi$ between the two arms,  and we measure the
intensity at the detector  as a function of $\chi$.} \label{exp_set}
\end{figure}

With the decomposition $A=UR$,  Eq. (\ref{rr}) and  Eq. (\ref{uu}), we now go to the measurement of non-Hermitian operator $A$. The schematic setup is illustrated in Fig. \ref{exp_set}. The average of $A$ in  state $|\phi_{in}\rangle$, $\langle\phi_{in}|A|\phi_{in}\rangle\equiv |A|e^{i\zeta}$ can be readout from the average intensity measured by the detector.
One point deserved to be addressed is that this set up can work for both polarized photons and two-level atoms. So, we would not specify which interferometer it is (Mach-Zehnder or Ramsey).
In bases spanned by $|a,1\rangle, |b,1\rangle,|a,2\rangle, |b,2\rangle$ ($|x,n\rangle\equiv|x\rangle\otimes|n\rangle, x=a,b; n=1,2$),
these operation $B_i (i=1 ,2), R, U^\dagger, e^{i\chi}$ can be written as

\begin{eqnarray}
B_{1,2}&=& \frac{1}{\sqrt{2}}\left(\begin{array}{cccc}
1 & 0 & i &0 \\
0 & 1 & 0 &i \\
i & 0 & 1 &0 \\
0 & i & 0 &1 \\
 \end{array}\right ),\quad
R= \left(\begin{array}{cccc}
R_{aa} & R_{ab} & 0 &0 \\
R_{ba} & R_{bb} & 0 &0 \\
0 & 0 & 1 &0 \\
0 & 0 & 0 &1 \\
 \end{array}\right),\nonumber\\
e^{i\chi}&=& \left(\begin{array}{cccc}
1 & 0 & 0 &0 \\
0 & 1 & 0 &0 \\
0 & 0 & e^{i\chi} &0 \\
0 & 0 & 0 &e^{i\chi} \\
 \end{array}\right ),\quad
U^\dagger= \left(\begin{array}{cccc}
1 &0 & 0 &0 \\
0 & 1 & 0 &0 \\
0 & 0 & U^\dagger_{aa} &U^\dagger_{ab}  \\
0 & 0 & U^\dagger_{ba} &U^\dagger_{bb} \\
 \end{array}\right),
\end{eqnarray}
where $R_{ab}=\langle a|R|b\rangle$, and similar denotations hold  for $R_{aa}, R_{bb}, R_{ba}$ and $U^\dagger_{xy}, x,y=a,b.$ With the same bases, the input state that only occupies spatial state $|1\rangle$ takes,
\begin{eqnarray}
|\phi_{in}\rangle&=& \left(\begin{array}{c}
\phi_a \\
\phi_b  \\
0 \\
0  \\
 \end{array}\right ), \phi_a\equiv \langle a|\phi_{in}\rangle, \, \phi_b\equiv \langle b|\phi_{in}\rangle.
\end{eqnarray}
The output state  reads
\begin{equation}
|\phi_{out}\rangle = B_2 e^{i\chi} U^\dagger R B_1|\phi_{in}\rangle,
\end{equation}
simple algebra yields,
\begin{eqnarray}
|\phi_{out}\rangle&=\frac 1 2 & \left(\begin{array}{c}
R_{aa}\phi_a+R_{ab}\phi_b-(U^\dagger_{aa}\phi_a+U^\dagger_{ab}\phi_b)e^{i\chi} \\
R_{ba}\phi_a+R_{bb}\phi_b-(U^\dagger_{ba}\phi_a+U^\dagger_{bb}\phi_b)e^{i\chi} \\
i[R_{aa}\phi_a+R_{ab}\phi_b+(U^\dagger_{aa}\phi_a+U^\dagger_{ab}\phi_b)e^{i\chi}] \\
i[R_{ba}\phi_a+R_{bb}\phi_b+(U^\dagger_{ba}\phi_a+U^\dagger_{bb}\phi_b)e^{i\chi}] \\
 \end{array}\right ).
\end{eqnarray}
The intensity the detector measures can be represented by
\begin{equation}
I(\chi)=|\langle\phi_D|\phi_D\rangle|^2, \quad |\phi_D\rangle\equiv D |\phi_{out}\rangle,
\end{equation}
with
\begin{equation}
D= \left(\begin{array}{cccc}
0 & 0 & 0 &0 \\
0 & 0 & 0 &0 \\
0 & 0 & 1 &0 \\
0 & 0 & 0 &1 \\
\end{array}\right ),
\end{equation}
leading to
\begin{equation}
I(\chi)=\frac 1 4 \left (\frac{}{}1+\langle\phi_{in}|R^2|\phi_{in}\rangle +2|\langle \phi_{in}|A|\phi_{in}\rangle|
\cos(\chi-\zeta)\right),
\end{equation}
where $\zeta$ is the argument of $\langle \phi_{in}|A|\phi_{in}\rangle$, i.e.,   $\zeta=\arg\langle \phi_{in}|A|\phi_{in}\rangle.$
In experiment, the intensity $I(\chi)$  together with $\langle\phi_{in}|R^2|\phi_{in}\rangle$ which is the average of Hermitian operator $R^2$  can  determine
the average of non-Hermitian operator $A,$ as both $|\langle \phi_{in}|A|\phi_{in}\rangle|$ and $\zeta$ can be inferred from the intensity $I(\chi)$.

\end{document}